\documentclass[aps,pra,reprint,floatfix,longbibliography]{revtex4-2}

\usepackage{microtype}
\usepackage{amsmath,amssymb,amsthm,mathtools}
\usepackage{bm}
\usepackage{booktabs}
\usepackage{array}
\usepackage{tabularx}
\usepackage{graphicx}
\usepackage{xcolor}
\usepackage[colorlinks=true,citecolor=blue!55!black,linkcolor=blue!55!black,urlcolor=blue!55!black]{hyperref}
\newcommand{\safegraphic}[2]{%
  \IfFileExists{#2}{\includegraphics[width=#1]{#2}}{%
    \fbox{\parbox[c][0.9in][c]{\dimexpr#1-2\fboxsep-2\fboxrule\relax}{%
      \centering\footnotesize Figure file not uploaded:\\
      \texttt{\detokenize{#2}}}}%
  }%
}
\newcommand{\cref}[1]{\autoref{#1}}
\newcommand{\Cref}[1]{\autoref{#1}}

\newcommand{\Tr}{\operatorname{Tr}}
\newcommand{\id}{\mathbb{I}}
\newcommand{\cH}{\mathcal{H}}
\newcommand{\cM}{\mathcal{M}}
\newcommand{\cE}{\mathcal{E}}
\newcommand{\cL}{\mathcal{L}}
\newcommand{\cP}{\mathcal{P}}
\newcommand{\cA}{\mathcal{A}}
\newcommand{\cX}{\mathcal{X}}
\newcommand{\Perr}{P_{\mathrm{err}}}
\newcommand{\Padstar}{P_{\mathrm{ad}}^{\star}}
\newcommand{\Pcohstar}{P_{\mathrm{coh}}^{\star}}
\newcommand{\Pcollstar}{P_{\mathrm{coll}}^{\star}}
\newcommand{\ppt}{\mathrm{PPT}}

\newcommand{\adapt}{\mathrm{ad}}

\newcommand{\Tpt}{\mathsf{T}_1}
\newcommand{\ket}[1]{\lvert #1\rangle}
\newcommand{\bra}[1]{\langle #1\rvert}

\newtheorem{definition}{Definition}
\newtheorem{theorem}{Theorem}
\newtheorem{proposition}{Proposition}
\newtheorem{lemma}{Lemma}
\newtheorem{corollary}{Corollary}
\theoremstyle{remark}
\newtheorem{remark}{Remark}

\begin{document}
\title{Certified Separation Between Coherent and Adaptive Temporal Processing of Quantum-Error-Correction Readout}


\author{Shreya Savadatti}
\email{shreya.savadatti@nlr.gov} 
\affiliation{Carnegie Mellon University, Pittsburgh, PA, 15213, USA}
\affiliation{Computational Science Center, National Laboratory of the Rockies, Golden, CO, 80401, USA}

\author{Matthew Reynolds}
\email{matthew.reynolds@nlr.gov}
\affiliation{Computational Science Center, National Laboratory of the Rockies, Golden, CO, 80401, USA}

\author{Jonathan Maack}
\affiliation{Computational Science Center, National Laboratory of the Rockies, Golden, CO, 80401, USA}

\author{Franz Franchetti}
\affiliation{Department of Electrical and Computer Engineering, Carnegie Mellon University, Pittsburgh, PA 15213, USA}

\author{Wesley Jones}
\affiliation{Computational Science Center, National Laboratory of the Rockies, Golden, CO, 80401, USA}


\begin{abstract}
Repeated quantum-error-correction readout is normally converted into a
classical time series: each syndrome pulse is measured on arrival and only the
resulting classical record is retained. Nonorthogonal readout signals raise
the question of whether preserving quantum information between rounds can
improve the final recovery decision. We formulate this comparison as a
loss-weighted state-discrimination problem. For an idealized two-round
benchmark with quadrature phase-shift-keying (QPSK) signals, a receiver with a
seven-dimensional persistent register has certified recovery-decision error
below \(24.2100\%\), whereas every separate adaptive receiver has error above
\(25.1431\%\). The certified separation is at least \(0.933\) percentage
points. The loss counts incorrect final error-mask decisions in the benchmark.
In the strict architecture, where the counted register is the entire
persistent state, four dimensions are necessary and sufficient to outperform
the adaptive class. Certified separations also occur at three and four rounds
with register dimensions \(8\) and \(36\), compared with \(16\) and \(64\)
for full-prefix storage. We relate memory dimension to the structure of the
temporal measurement, give a sufficient memory bound for collective
measurements, and quantify robustness to register dephasing and accumulated
channel error.
\end{abstract}
\maketitle

\section{Introduction}

Repeated quantum error correction (QEC) is inherently a temporal inference
problem.  A single syndrome observation can be ambiguous because data faults
and readout noise both affect the observed record, so practical decoders
combine information across multiple extraction rounds.  Recent work has further
shown that retaining analog or soft readout information can improve on hard
syndrome decisions \citep{pattison2021soft,ali2024analog,hanisch2026soft}.
These results make clear that not only the syndrome record, but also how that
record is represented and retained, can affect the final recovery decision.

In the usual measurement-based architecture, each syndrome extraction produces
a physical readout signal that is measured as it arrives.  The resulting
classical record can be retained in considerable detail: a decoder may keep
analog values, assign likelihoods, model correlations across rounds, and use
the complete history when selecting a recovery.  Such processing can preserve
much more information than a hard syndrome bit.  It nevertheless begins after
the arriving quantum signal has been measured.  By the time temporal decoding
takes place, the information available to the decoder is classical.

At the other end of the design space, coherent and measurement-free QEC
architectures avoid the conventional measurement-and-feedback loop by retaining
quantum information within a coherent controller or by implementing correction
through autonomous dynamics
\citep{kerckhoff2009embedded,crow2016measurementfree,butt2025measurementfree,
butt2026demonstration}.  These approaches show that syndrome information need
not always be converted immediately into a classical record, but they also
change how error correction itself is implemented.

Here we consider an intermediate possibility.  Before measurement, the
finite-strength signals associated with distinct syndromes need not be
orthogonal, so there is an operational distinction between remembering the past
\emph{after} measurement and preserving part of the past \emph{before}
measurement.  We leave the syndrome-extraction process and the emitted readout
signals unchanged, but alter when those signals must become classical.  Rather
than measuring every signal separately as it arrives, a finite quantum memory
may retain information from earlier signals and interact with later ones,
allowing some or all measurement to be deferred until later signals have
arrived.

We therefore ask whether, for the same sequential readout ensemble and
recovery objective, preserving quantum information across rounds can outperform
every receiver that measures each signal separately and retains only classical
information between rounds.

Answering this question requires more than comparing against a particular
decoder.  Failure of one classical algorithm could reflect the choice of
decoder rather than a limitation imposed by measurement itself.  We therefore
define a deliberately broad separate adaptive class: each arriving signal may
undergo an arbitrary quantum instrument, the receiver may retain an
unrestricted classical transcript, and all later instruments may depend on that
transcript.  The only excluded resource is a quantum state carrying information
from an earlier signal into a later round.

Direct optimization over this adaptive class is difficult.  Every such
receiver induces measurement operators that are separable across the temporal
signal factors and therefore positive under partial transpose (PPT).  This
containment gives a tractable semidefinite-programming relaxation:
a lower bound obtained over the larger PPT class is automatically a lower bound
on the error attainable by any receiver in the separate adaptive class.
Consequently, a feasible coherent receiver whose error lies below this bound is
separated from the entire measure-and-feed-forward class, rather than from a
particular decoder found numerically.

For the two-round QPSK benchmark studied here, a \(K=7\) streaming receiver
achieves a recovery-decision error of approximately \(0.2421\), while the
certified adaptive lower bound is approximately \(0.2514\).  The resulting
separation is about \(9.3\times10^{-3}\), or \(0.93\) percentage points; the
outward-rounded values are stated in Theorem~\ref{thm:main}.  This gap
certifies a separation between the two receiver classes under the same signal
ensemble and recovery objective.

We next determine the persistent-memory requirements of this separation.  We
distinguish the broad coherent class, which may retain classical side
information, from a strict fixed-memory architecture in which the counted
register is the only persistent state.  For the latter model, the two-round
benchmark has the exact threshold \(K_{\rm sep}=4\), while
\(4\le K_{\rm coll}\le7\).  For each explicit receiver, the certified
advantage persists up to a rigorously bounded dephasing strength on the
persistent register.  More generally, the entanglement-breaking analysis shows
that any receiver exhibiting such an advantage must have at least one
persistent boundary whose complete classically flagged instrument is not
measure-and-prepare.  The
temporal-separability argument extends to arbitrary depth, and certified
separations are obtained with \(K=8<16\) at \(W=3\) and \(K=36<64\) at
\(W=4\).  These finite-depth results establish persistence of the effect
beyond two rounds; asymptotic memory scaling remains open.

We also derive a diamond-norm accumulation bound that gives sufficient
implementation-error budgets for the explicit streaming receivers.  Separately,
a stored-pulse model translates phase retention into approximate
storage-lifetime requirements and compares their scale with experimentally
reported component coherence times.  This comparison provides a
component-level feasibility check.  The
architecture also changes where latency enters the recovery pipeline:
information that would otherwise be converted into a classical record between
rounds may instead be carried forward coherently, replacing some inter-round
measurement and feed-forward requirements with coherent storage and control.
This suggests a latency-coherence tradeoff whose quantitative
characterization requires a hardware-specific timing model.

\begin{figure*}[t]
\centering
\safegraphic{0.98\textwidth}{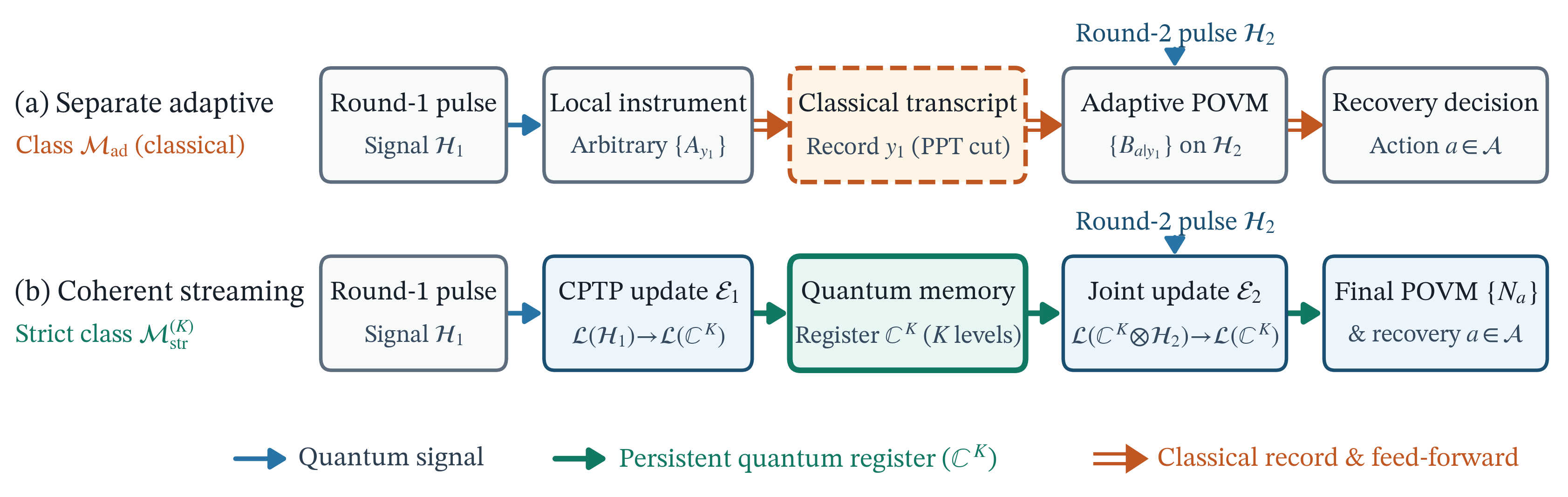}
\caption{\textbf{Receiver-class comparison.} Both receivers receive the same
pulse ensemble, prior, clock, and recovery objective.  The separate adaptive
receiver may preserve an arbitrarily rich classical record.  The coherent
receiver additionally preserves a finite-dimensional quantum system between
rounds. In the strict memory-\(K\) model, the labeled \(K\)-dimensional register
is the only persistent state; no additional persistent classical side register
is available.}
\label{fig:architecture}
\end{figure*}

\section{Related work}

\paragraph{Soft syndrome decoding.}
Experimental readout produces analog voltages or photon counts before
thresholding, and a hard decision discards information about the confidence
of that assignment. Soft and analog decoders retain part of this
information and can lower logical error relative to hard-decision decoding
\citep{pattison2021soft,ali2024analog,hanisch2026soft}.
These results establish that the representation of readout information
affects QEC performance even when the syndrome-extraction circuit remains
unchanged. However, soft decoding operates on information that has already
been converted into a classical measurement record. Our work asks whether
the measurement itself imposes a fundamental limitation, even when all
available classical information is retained. To isolate this distinction,
our separate adaptive class permits arbitrary local measurements,
continuous outcomes, an unrestricted classical transcript, and unrestricted
classical post-processing. We establish a certified separation from this entire class, demonstrating that, for the benchmark considered, coherent processing achieves a lower recovery error than any separate adaptive receiver, even with unrestricted classical memory and post-processing.

\paragraph{Coherent and measurement-free QEC.}
Coherent feedback and measurement-free error correction avoid the standard
mid-circuit measurement loop
\citep{kerckhoff2009embedded,crow2016measurementfree,
butt2025measurementfree,brechtelsbauer2025biased,butt2026demonstration}.
In these architectures, a coherent controller may participate directly
in syndrome extraction, feedback, or autonomous correction. Recent theory
provides scalable fault-tolerant and noise-biased measurement-free
constructions
\citep{butt2025measurementfree,brechtelsbauer2025biased}, while trapped-ion
experiments have demonstrated measurement-free logical operations
\citep{butt2026demonstration}.
A separate spatio-temporal framework describes adaptive error correction
using sequential interventions \citep{tanggara2024strategic}.
Our work addresses a different question. Rather than modifying syndrome
extraction or implementing coherent correction, we hold the extraction
circuit, emitted pulse ensemble, trajectory prior, and recovery loss fixed.
Only the processing of the emitted temporal signals changes: each pulse
is either measured on arrival, with its outcome retained classically, or
processed through a persistent quantum memory. This isolates the value of
quantum memory at the receiver level and establishes a certified advantage
without attributing the improvement to changes in the underlying
error-correction circuit.

\paragraph{Restricted quantum measurements.}
Minimum-error discrimination admits a primal-dual semidefinite programming
(SDP) formulation
\citep{helstrom1976detection,yuen1975optimum,watrous2018theory}.
For multipartite measurements, local operations and classical communication
(LOCC), separable measurements, and positive-partial-transpose (PPT)
measurements form a useful hierarchy
\citep{matthews2009distinguishability,peres1996separability,
horodecki1996separable,cosentino2013ppt}.
PPT constraints provide a tractable outer relaxation of separable
measurements. We apply this hierarchy to a temporal discrimination problem
in which the tensor factors represent readout signals arriving in
successive rounds. Measuring each signal and retaining only its classical
outcome produces adaptive-tree effects that are separable and therefore
PPT. The resulting SDP consequently bounds the entire defined
measure-and-feed-forward class, rather than a particular decoder or
measurement policy. Our contribution is to combine this relaxation with
a coherent receiver construction and rigorous verification of the
optimization bounds. The resulting certificate establishes a strict
separation between the receiver classes, rather than relying on numerical
comparisons between individual measurement strategies.

\paragraph{Quantum memories and entanglement-breaking channels.}
Entanglement-breaking (EB) channels are precisely measure-and-prepare
channels \citep{horodecki2003eb}. Resource-theoretic approaches have used
discrimination tasks to verify whether a quantum memory is not
entanglement-breaking \citep{rosset2018memory}. However, the temporal
interpretation of this property requires care: EB channels can still
support nonclassical multitime correlations under broader operational
notions of memory \citep{vieira2025eb}.
Our result is therefore specific to the causal receiver architecture
considered here. If the complete classically flagged instrument at each persistent boundary
is measure-and-prepare, the resulting statistics can be reproduced by a
separate adaptive receiver. Consequently, a receiver that achieves the
certified separation must have at least one boundary that is not
measure-and-prepare. This gives the separation a resource interpretation:
the advantage requires genuinely quantum memory, not merely a different
classical encoding of earlier measurement outcomes. The simulability
statement does not extend to arbitrary multitime intervention models.

\paragraph{Collective and sequential quantum receivers.}
Collective optical receivers can outperform symbol-by-symbol detection
on structured coherent-state ensembles
\citep{guha2011structured,barnett2009discrimination}.
Such improvements arise because joint measurements need not reduce
each signal to a classical outcome before subsequent signals are
processed. Previous work has also shown that small persistent quantum
memories can reproduce collective discrimination performance in
streaming settings
\citep{blumekohout2012ideal,klimek2016memory}.
Our work builds on this connection but considers a recovery-weighted
objective and compares finite-memory coherent receivers against the
entire class of arbitrary local measurements with classical memory
and feed-forward. A finite persistent register restricts the
operator-space complexity of the induced temporal effects. In the
strict architecture considered here, this appears as an
operator-Schmidt-rank bound across each chronological cut, corresponding
to the matrix-product-operator bond-dimension constraint associated
with finite sequential memory
\citep{verstraete2004mpdo,schollwock2011dmrg}.
We use these structural constraints to establish lower bounds on the
required memory dimension and combine them with certified receiver
constructions. For the two-round benchmark, this determines the exact
minimum memory dimension needed to outperform every separate adaptive
receiver, strengthening the connection between sequential
discrimination performance and the quantum resources required to
achieve it.

\paragraph{Quantum reservoir computing.}
Quantum reservoir computing (QRC) uses a persistent dynamical system
and a simple trained readout to process temporal data
\citep{fujii2017qrc,tanaka2019physical}.
Its use of persistent quantum dynamics makes it a natural candidate
for processing temporally correlated readout signals without measuring
each signal independently. Appendix~\ref{app:qrc} examines this
connection for the \(W=3\) benchmark by comparing a homogeneous
reservoir with a clocked construction. The clocked construction achieves
lower error in all ten paired best-of-three comparisons with the
matched homogeneous reservoir. Because the clocked construction was
informed by the optimized channel, we treat these results as
exploratory rather than as evidence of a general architectural
advantage. They illustrate how the receiver framework may inform
practical sequential architectures, while the formal contribution
of this paper remains the certified separation and the associated
memory-resource bounds.

\section{Physical background and notation}
\label{sec:background}

\subsection{From physical errors to readout signals}

A physical qubit is an individual two-level quantum system. A logical qubit
stores quantum information in a subspace of several physical qubits, so that
specified errors can be detected without measuring the unknown logical
amplitudes. In a stabilizer code, commuting check operators distinguish error
sectors. Their outcomes form the \emph{syndrome}: information about an error,
rather than a measurement of the stored logical state. An ancillary qubit is
a helper system coupled to the data to extract a check outcome, then read
out and reset.

For example, the three-qubit repetition encoding
\(\ket{0_L}=\ket{000}\), \(\ket{1_L}=\ket{111}\) protects against a single
bit flip, rather than arbitrary quantum noise; phase errors require
additional protection. Here \(\ket0,\ket1\) denote computational-basis states,
\(X\) exchanges them, and \(Z=\ket0\!\bra0-\ket1\!\bra1\).
The parity checks \(Z_1Z_2\) and \(Z_2Z_3\) give syndrome bits
\(b_1\oplus b_2\) and \(b_2\oplus b_3\) for error mask
\(b=(b_1,b_2,b_3)\), where \(\oplus\) is addition modulo two.
This convention illustrates syndrome extraction; the phase lookup used in
the certified benchmark is a separate part of its specification.

In dispersive microwave readout, an ancilla changes the response of a
resonator and thereby the state of an outgoing pulse \citep{blais2021cqed}.
Finite pulses associated with different states can overlap, so a measurement
need not reveal their labels without error. Repeated extraction rounds help
distinguish ambiguous observations and track faults occurring between rounds.
A receiver combines the readout signals to select a correction, either
applied physically or recorded in a \emph{Pauli frame}, a classical record of
corrections used when interpreting later operations and measurements.

\begin{figure*}[t]
\centering
\safegraphic{0.64\textwidth}{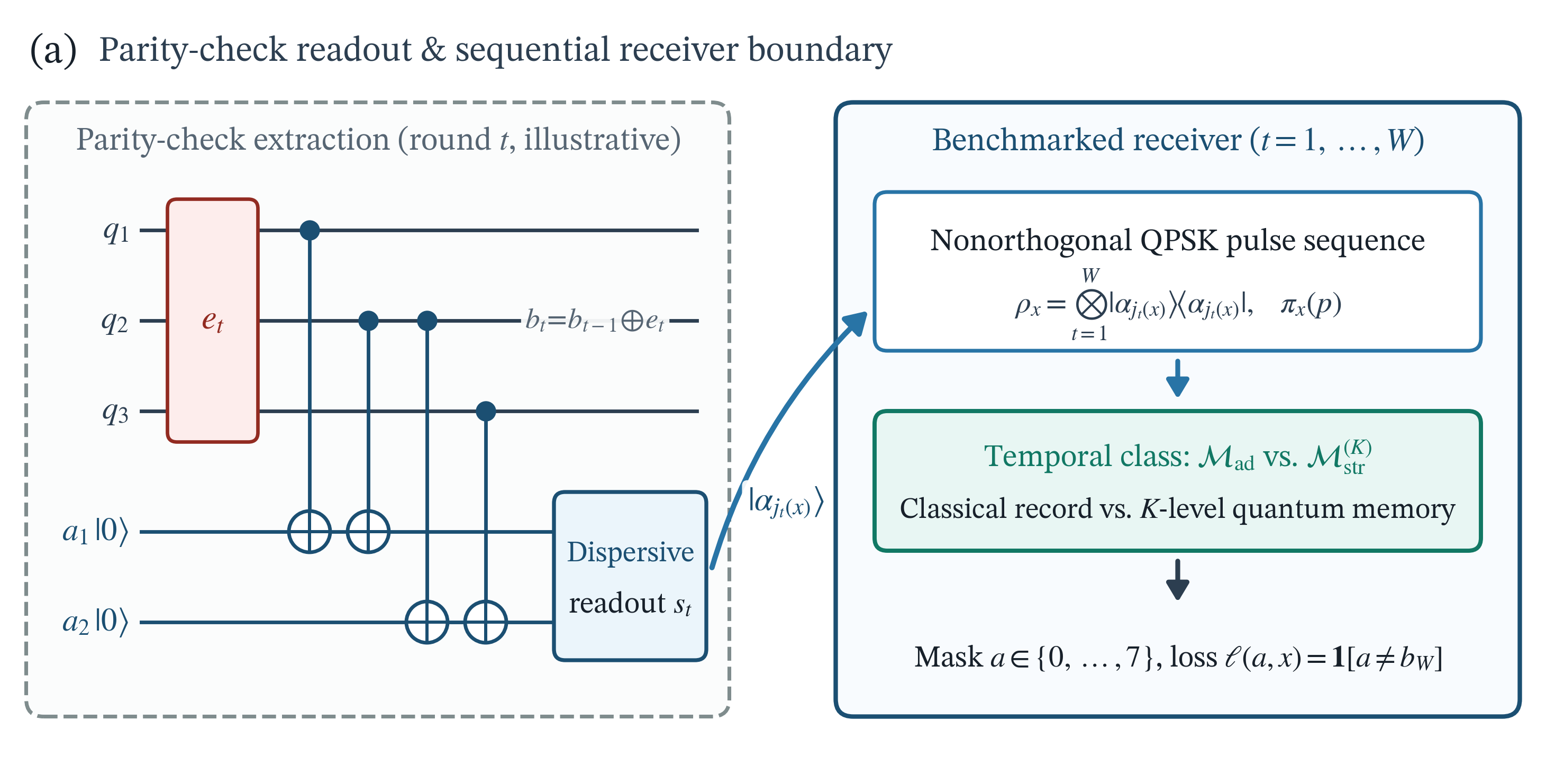}\hfill
\safegraphic{0.32\textwidth}{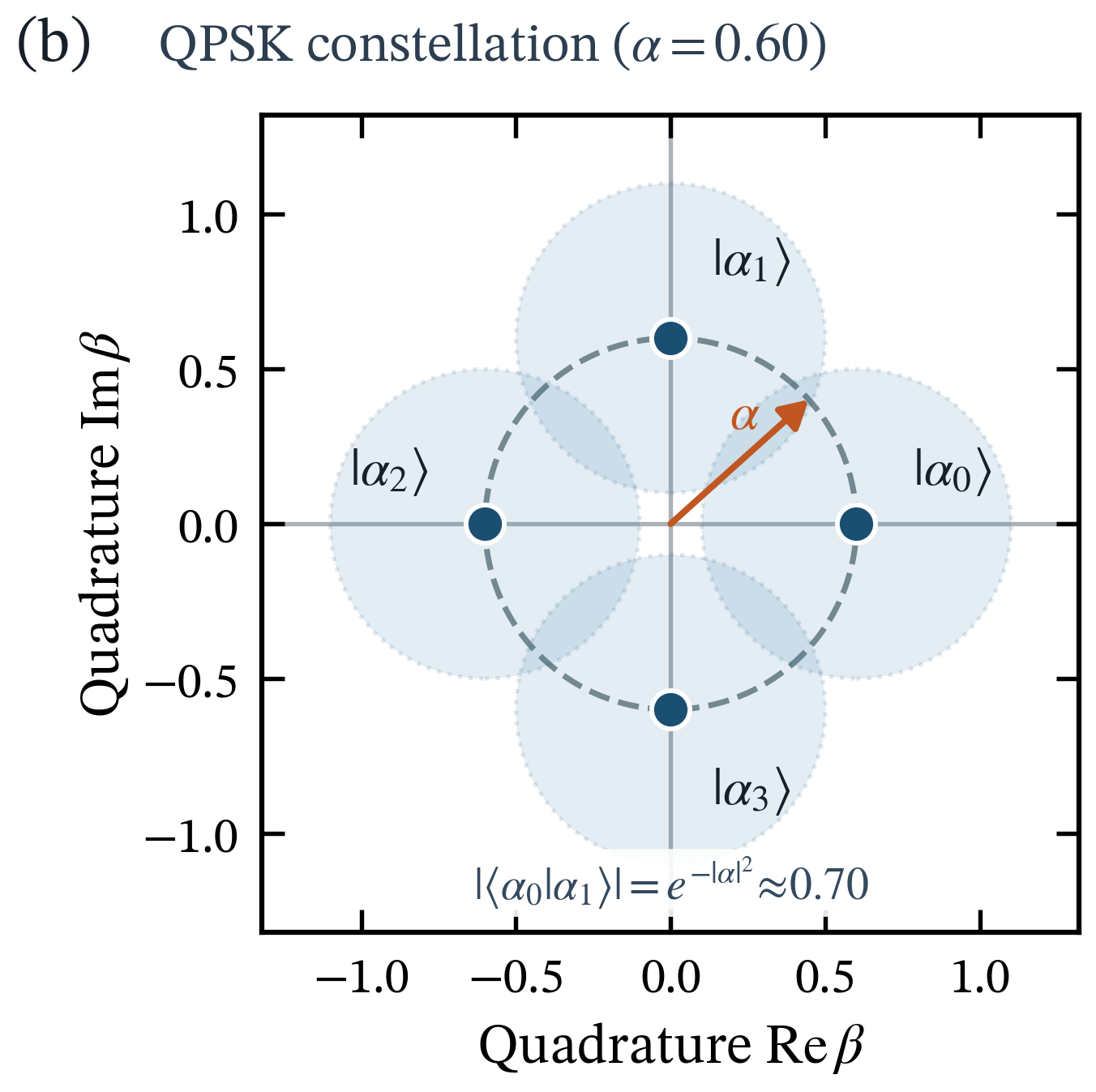}
\caption{\textbf{Readout generation and signal encoding.} Left: a conceptual
parity-check extraction circuit couples data qubits to ancillary systems and
generates readout signals. The receiver processes these signals and selects a
recovery; the extraction circuit is illustrative rather than simulated in the
benchmark. Right: quadrature phase-shift keying (QPSK) places four
coherent-state amplitudes at phases \(0,\pi/2,\pi,3\pi/2\), with radius
\(\alpha=0.60\). The axes are real and imaginary field amplitude. Distinct
points represent nonorthogonal quantum states. The phase index \(j\) is
assigned to a syndrome by the benchmark lookup, not by the drawing.}
\label{fig:physical-background}
\end{figure*}

The benchmark below models these emitted signals directly. Each two-bit
syndrome is assigned to one of four coherent pulse states. The ancilla circuit
is physical motivation rather than a simulated component of the numerical
model. The question is whether a receiver can use the same pulse sequence
more effectively by retaining quantum information between arrivals.
Figure~\ref{fig:physical-background} shows this connection.

\subsection{Mathematical description of signals and receivers}

A Hilbert space \(\cH\) describes a quantum system, and \(\cL(\cH)\) is the
space of linear operators on it. The set \(\mathcal D(\cH)\) consists of
density operators: positive semidefinite operators of trace one. We write
\(A\succeq0\) when \(\bra v A\ket v\ge0\) for every vector \(\ket v\), and
\(A\succeq B\) when \(A-B\succeq0\). The trace \(\Tr A\) is the sum of
diagonal entries, \(A^\dagger\) is the Hermitian adjoint,
\(A^{\mathsf T}\) is transpose in the stated basis, and \(\id\) is the
identity operator. A ket \(\ket\psi\) denotes a state vector and
\(\ket\psi\!\bra\psi\) its pure-state density operator. Tensor products,
written \(\otimes\), describe joint systems; here their factors label
different temporal signals. A partial trace \(\Tr_B\) discards subsystem
\(B\), while \(A^{\mathsf T_S}\) transposes only the factors indexed by \(S\).

A positive operator-valued measure (POVM) \(\{M_a\}\) is a general quantum
measurement: its positive effects sum to the identity, and
\(\Tr(\rho M_a)\) is the probability of decision \(a\) for input \(\rho\).
A quantum instrument also specifies the state update associated with each
outcome. A completely positive trace-preserving (CPTP) map is a quantum
channel: it maps valid states to valid states, including when applied to part
of a larger system. An entanglement-breaking (EB) channel measures its input
and prepares a new state depending only on the outcome.

We use \(W\) for the number of readout rounds and \(K\) for the Hilbert-space
dimension of the persistent receiver register. Thus \(K\) describes the size
of the quantum state available from one round to the next, rather than a
physical-qubit count. It embeds in \(\lceil\log_2K\rceil\) ideal qubits,
where \(\lceil\cdot\rceil\) denotes the ceiling function.
The single-signal dimension is \(d_{\mathrm{sig}}=\dim\cH_t\) when common
to all rounds. The separate symbol \(d_{\mathrm{code}}\) denotes code
distance, the minimum weight of a nontrivial logical Pauli operator in a
stabilizer code.

\paragraph{Operator norms and measurement bounds.}
The receiver comparison uses matrix constraints and norms to bound
measurement performance and sensitivity to perturbations.
For operators, \(\lVert A\rVert_1=\Tr\sqrt{A^\dagger A}\) is the trace norm,
\(\lVert A\rVert_\infty\) the largest singular value, and
\(\lVert A\rVert_F=\sqrt{\Tr(A^\dagger A)}\) the Frobenius norm.
An effect is separable across temporal factors when it is a limit of sums of
tensor products of positive effects. Such effects have positive partial
transpose (PPT), the necessary condition used to bound adaptive receivers.
A semidefinite program (SDP) optimizes a linear objective with positive-matrix
constraints; a feasible dual witness supplies a bound on every primal
solution.

\begin{table}[t]
\centering
\small
\caption{Main quantities and their physical meanings.}
\label{tab:notation}
\begin{tabular}{@{}ll@{}}
\toprule
Symbol & Meaning\\
\midrule
\(x,\pi_x\) & \parbox[t]{0.60\columnwidth}{Error trajectory and its probability}\\[3pt]
\(\rho_x\) & \parbox[t]{0.60\columnwidth}{Readout state emitted for that trajectory}\\[3pt]
\(a,M_a\) & \parbox[t]{0.60\columnwidth}{Recovery action and its measurement effect}\\[3pt]
\(\ell(a,x)\) & \parbox[t]{0.60\columnwidth}{Loss for that action on trajectory \(x\)}\\[3pt]
\(C_a,S_a\) & \parbox[t]{0.60\columnwidth}{Cost and success operators weighted by the prior}\\[3pt]
\(r(M;C)\) & \parbox[t]{0.60\columnwidth}{Average loss of a specified receiver}\\[3pt]
\(W,K\) & \parbox[t]{0.60\columnwidth}{Number of rounds and persistent-register dimension}\\[3pt]
\(d_{\rm sig},d_{\rm code}\) & \parbox[t]{0.60\columnwidth}{Signal dimension and code distance}\\[3pt]
\(P^\star,K_{\rm sep}\) & \parbox[t]{0.60\columnwidth}{Class-optimal risk and minimum strict-register dimension that outperforms the adaptive class}\\
\bottomrule
\end{tabular}
\end{table}

\section{Sequential recovery problem and receiver classes}

\subsection{Loss-weighted pulse ensemble}

Let \(W\) denote the number of readout rounds.  A latent trajectory
\(x\in\cX\) specifies the underlying fault history together with the syndrome
labels it produces over those rounds.  The trajectory is assigned prior
probability \(\pi_x\) and generates an arbitrary receiver-independent temporal
state
\begin{equation}
 \rho_x\in\mathcal D(\cH_{1:W}),\qquad
 \cH_{1:W}:=\bigotimes_{t=1}^{W}\cH_t .
 \label{eq:trajectory}
\end{equation}
Thus \(\rho_x\) may be mixed and may contain arbitrary classical or quantum
temporal correlations.  Temporal separability below is a property of the
receiver protocol, not of this input state.  The only source assumption is that
the joint state is fixed independently of receiver actions.
If an earlier receiver intervention causally changes later emissions, a process
tensor or quantum-comb description is required instead
\citep{pollock2018process,white2022process}.  The present benchmark has no such
feedback into the
source: syndrome extraction and the emitted ensemble are held fixed.

Let \(a\in\cA\) label a recovery action and
\(\ell(a,x)\in[0,1]\) its recovery loss.  Operationally, \(a\) can represent a
physical correction or a Pauli-frame update. In the binary-loss benchmark
below, the loss is zero exactly when the selected recovery mask equals the
final physical-error mask, and one otherwise. Code-level logical success can
identify different physical errors as equivalent; the benchmark instead uses
exact mask matching. Define
\begin{equation}
 C_a:=\sum_{x\in\cX}\pi_x\,\ell(a,x)\rho_x .
 \label{eq:cost}
\end{equation}
For an effective POVM \(\{M_a\}_{a\in\cA}\), the average recovery risk is the
linear functional
\begin{equation}
 r(M;C):=\sum_{a\in\cA}\Tr(C_aM_a).
 \label{eq:risk}
\end{equation}
This expression averages the loss over the possible trajectories and the
receiver's decisions. For the benchmark below, it is the probability of an
incorrect final-mask decision, called the recovery-decision error.
We distinguish this risk of a specified receiver from the class optima,
\begin{equation}
\begin{aligned}
 \Padstar(C)&:=\inf_{M\in\cM_{\rm ad}}r(M;C),\\
 \Pcohstar(C)&:=\inf_{M\in\cM_{\rm coh}}r(M;C).
\end{aligned}
 \label{eq:class-optima}
\end{equation}
Here \(\cM_{\rm ad}\) and \(\cM_{\rm coh}\) denote the receiver classes defined
below; \(r(M;C)\) is the risk of a specified receiver, whereas starred
quantities denote class optima.
Each \(C_a\) incorporates the trajectory prior, emitted temporal state, and
recovery loss associated with recovery action \(a\).  The resulting receiver
optimization is a linear quantum decision problem.  The objective rewards
successful recovery rather than exact reconstruction
of the syndrome history.

\subsection{Two-round benchmark}
\label{sec:benchmark}

The numerical instance specializes this construction to two sequential readout
rounds.  In each round the relevant syndrome has four possible values,
represented by two syndrome bits.  A fixed lookup maps these four labels
one-to-one to a QPSK phase \(j\in\{0,1,2,3\}\), producing the coherent state
\begin{equation}
 \ket{\alpha_j}=\ket{\alpha e^{i\pi j/2}},\qquad \alpha=0.60 .
 \label{eq:qpsk}
\end{equation}
The amplitude \(\alpha\) is dimensionless, with mean photon number
\(|\alpha|^2=0.36\). Four equally spaced phases encode the four syndrome
labels. Distinct amplitudes in phase space need not be orthogonal states.
Coherent pointer states arise naturally in dispersive microwave readout, where
finite-amplitude responses associated with different system states are not
perfectly distinguishable \citep{blais2021cqed}.  The QPSK encoding used here
provides a symmetric four-state benchmark of this nonorthogonal-readout
setting.
At \(\alpha=0.60\) the four pulses are nonorthogonal, so no measurement can
identify all four labels perfectly.  We set \(p=0.30\) in the trajectory
prior \(\pi_x(p)\).  The final decision set contains eight recovery masks.
The loss and trajectory construction are specified below and in
Appendix~\ref{app:benchmark}.

\paragraph{Error trajectories and the recovery criterion.}
The per-round error increment is \(e_t\in\{0,1,2,4\}\), where each nonzero
integer mask flips one of the three physical bits. Increments are independent,
with probabilities \(1-p,p/3,p/3,p/3\). The cumulative error is the bitwise
exclusive OR \(b_t=e_1\oplus\cdots\oplus e_t\), and
\(x=(e_1,\ldots,e_W)\) specifies the full trajectory. The syndrome of the
cumulative mask is \(s(b_t)\), giving phase index \(j_t(x)=j(s(b_t))\).
The eight actions \(a\in\{0,\ldots,7\}\) are all three-bit recovery masks,
and the loss is
\begin{equation}
 \ell(a,x)=\mathbf 1[a\ne b_W],
 \label{eq:mask-loss}
\end{equation}
where the indicator \(\mathbf 1[\cdot]\) is one if its condition holds and
zero otherwise. An error near \(24.21\%\) therefore means an incorrect final
mask on that fraction of the ensemble.

Exact mask matching cancels all modeled bit flips. For an encoded state, a
nonzero residual mask may instead leave a detectable error or induce a
logical operation. Counting every such residual as a failure is stricter than
identifying only logical equivalence classes. The reported numbers measure
this specified recovery decision, rather than a circuit-level logical-failure
rate.

\begin{table}[t]
\centering
\small
\caption{Per-round benchmark increments. The syndrome and pulse columns
describe a single increment starting from zero error. At later rounds the
lookup is applied to the cumulative mask \(b_t\). The symbolic functions
\(s\) and \(j\) denote the benchmark's syndrome and phase lookups.}
\label{tab:benchmark-map}
\begin{tabular}{@{}llll@{}}
\toprule
Increment & Probability & Syndrome & Pulse\\
\midrule
No flip, \(0\) & \(1-p=0.70\) & \(s(0)\) & \(\ket{\alpha_{j(s(0))}}\)\\
First bit, \(1\) & \(p/3=0.10\) & \(s(1)\) & \(\ket{\alpha_{j(s(1))}}\)\\
Second bit, \(2\) & \(p/3=0.10\) & \(s(2)\) & \(\ket{\alpha_{j(s(2))}}\)\\
Third bit, \(4\) & \(p/3=0.10\) & \(s(4)\) & \(\ket{\alpha_{j(s(4))}}\)\\
\bottomrule
\end{tabular}
\end{table}

For example, \(x=(1,0)\) is a first-bit flip in round one and no new flip
in round two, with probability \((p/3)(1-p)=0.07\).
The cumulative mask is one in both rounds, so both emitted pulses are
\(\ket{\alpha_{j(s(1))}}\). In the illustrative check convention of
Sec.~\ref{sec:background}, the syndrome is \(10\). Choosing \(a=1\)
has zero loss; every other mask has unit loss. The receiver sees only the
readout signals and must infer the recovery without knowing \(x\).

Unlike the general formulation in \cref{eq:trajectory}, this specific QPSK
benchmark uses conditionally independent emitted modes,
\begin{equation}
 \rho_x^{\mathrm{QPSK}}=
 \bigotimes_{t=1}^{W}
 \ket{\alpha_{j_t(x)}}\!\bra{\alpha_{j_t(x)}} .
 \label{eq:qpsk-product}
\end{equation}
The product form is therefore a specialization of the numerical instance, not
an assumption of the receiver-class theorem.

Although the pulse modes have an infinite-dimensional Fock space, the four
states used in each round span a four-dimensional subspace
\(\mathcal S_t\subset\cH_t\). Let \(P_t\) project onto that subspace and
write \(P=\bigotimes_t P_t\). Every benchmark state satisfies
\(P\rho_xP=\rho_x\). For any receiver effect \(M_a\), the compression
\(PM_aP\), regarded as an operator on \(\bigotimes_t\mathcal S_t\),
therefore gives the same outcome probabilities and risk. The compressed
effects remain a POVM on that space. For a separate adaptive receiver,
compression acts on each local factor of every product effect, preserving
separability and PPT. Its conditional local measurements remain complete on
\(\mathcal S_t\). The coherent update maps can likewise be restricted to
these supported signal inputs without changing their statistics. Thus the
finite-dimensional receiver results apply exactly to this ensemble, without
a photon-number cutoff.

We represent this finite-dimensional signal space directly through the
coherent-state Gram matrix
\begin{equation}
 G_{jk}=\langle\alpha_j|\alpha_k\rangle
 =\exp\!\left[-\frac{|\alpha_j|^2+|\alpha_k|^2}{2}
 +\alpha_j^*\alpha_k\right].
 \label{eq:gram}
\end{equation}
Here \(G_{jk}\) is the overlap of pulses \(j\) and \(k\), and the asterisk
denotes complex conjugation. A factorization \(G=V^\dagger V\) supplies coordinate vectors for the states
without truncating their photon-number expansion. The four states span
\(d_{\mathrm{sig}}=4\) dimensions; this is an effective representation of
the chosen ensemble, not a truncation of the electromagnetic field's full
Hilbert space. The certified calculation
reconstructs these coordinates and the cost operators directly from the
analytic Gram data using Arb/Acb ball arithmetic. No photon-number cutoff
enters the proof. Appendix~\ref{app:benchmark} describes the analytic
reconstruction and an independent truncated-Fock convergence check.
The parameter \(p\) is a benchmark trajectory parameter rather than a
calibrated device error rate. The operating point is chosen to provide a
nontrivial temporal-discrimination regime, rather than to model a
fault-tolerant device setting.

\subsection{Receiver classes}

The separate adaptive receiver carries an unrestricted classical transcript
between rounds. The general coherent receiver can retain quantum memory
alongside that transcript. The strict receiver permits only the counted
\(K\)-dimensional register. These are constraints on retained information,
rather than on classical computational power.

\begin{definition}[Separate adaptive receiver]
\label{def:adaptive}
At round \(t\), the receiver applies an arbitrary quantum instrument to the
current pulse.  Its choice may depend on the clock and the entire earlier
classical transcript.  It may use arbitrary local ancillas, randomness,
continuous outcomes, and unrestricted classical computation.  No quantum
state carried between rounds may depend on an earlier pulse; all information
retained from previous rounds is contained in the classical transcript.
\end{definition}

Thus a separate adaptive receiver may remember arbitrarily much about the
past, including a complete analog record, but every persistent record must be
classical.

For \(W=2\), coarse-graining the complete transcript into the final action
gives an effect
\begin{equation}
 M_a^{\adapt}=\int A_y\otimes B_{a|y}\,d\mu(y),
 \label{eq:adaptive}
\end{equation}
where \(A_y\succeq0\) is the first-round effect and
\(\{B_{a|y}\}_a\) is the conditional second-round POVM. Each
\(M_a^{\adapt}\) is therefore separable across the temporal cut.

\begin{definition}[Coherent receiver]
\label{def:coherent}
A coherent receiver is a causal receiver that may retain a finite-dimensional
quantum system between rounds. It may also retain an unrestricted classical
transcript and condition later operations on that transcript. The union over
all finite quantum-memory dimensions is denoted \(\cM_{\rm coh}\).
\end{definition}

\begin{definition}[Strict memory-\(K\) streaming receiver]
\label{def:strict-streaming}
A strict memory-\(K\) receiver consists of completely positive
trace-preserving maps
\begin{align}
 \cE_1&:\cL(\cH_1)\rightarrow\cL(\mathbb C^K),\nonumber\\
 \cE_t&:\cL(\mathbb C^K\otimes\cH_t)\rightarrow\cL(\mathbb C^K),
 \quad 2\le t\le W ,
 \label{eq:streammaps}
\end{align}
followed by a final memory POVM \(\{N_a\}\).
The \(K\)-dimensional register is the only persistent state between rounds:
no additional persistent classical record, quantum pointer, retained signal,
or entangled ancillary system is available outside it. Fresh within-round
ancillas may be introduced and discarded. The resulting receiver class is
denoted \(\cM_{\rm str}^{(K)}\).
\end{definition}

The memory-dimension results use \(\cM_{\rm str}^{(K)}\).  The receiver-class
separation instead compares \(\cM_{\rm ad}\) with the broader class
\(\cM_{\rm coh}\), which permits additional classical side information.
In \(\cM_{\rm str}^{(K)}\), the final POVM acts on the same counted
\(K\)-dimensional register.

The unrestricted collective receiver, which measures
\(\cH_{1:W}\) directly, supplies a global comparison point.  The streaming
model asks how much of that performance can be retained without storing the
entire pulse block.

\subsection{Arbitrary-depth structure of separate adaptive receivers}

\begin{theorem}[Temporal separability of classical-memory receivers]
\label{thm:temporal-separability}
Let \(\cH_1,\ldots,\cH_W\) be finite-dimensional temporal signal systems.
Every receiver that measures each signal locally, retains only classical
information between rounds, and classically maps its complete transcript to a
final action induces effects of the form
\begin{equation}
\begin{aligned}
 M_a=\int_{\Gamma_a} &E^{(1)}_{y_1}\otimes
 E^{(2)}_{y_2|y_1}\otimes\cdots\\
 &\otimes E^{(W)}_{y_W|y_{<W}}\,d\mu(y),
\end{aligned}
\label{eq:adaptive-tree-effect}
\end{equation}
where every factor is positive semidefinite.  Thus \(M_a\) is a positive conic
integral of product effects, equivalently a limit of finite positive sums of
product effects, and is fully separable across
\(\cH_1\otimes\cdots\otimes\cH_W\).  Consequently,
\begin{equation}
 M_a^{\mathsf T_S}\succeq0
 \quad\text{for every }S\subseteq\{1,\ldots,W\}.
 \label{eq:all-cuts-ppt}
\end{equation}
The receiver may use arbitrary discrete or continuous outcomes, private
randomness, unrestricted classical computation, and fresh or resettable local
ancillas. Outcome spaces are taken to be standard Borel, with measurable
conditional measurement kernels.
\end{theorem}
The proof is given in Appendix~\ref{app:temporal-proof}.

\begin{corollary}[Multipartite PPT outer bound]
\label{cor:multippt}
For any \(W\)-round recovery problem, minimizing the recovery loss over a
measurement class constrained to be PPT across any selected temporal cuts
gives a lower bound on the error of every separate adaptive receiver. If the
risk of a feasible coherent receiver is below a certified dual lower bound for
that relaxation, then it outperforms the entire separate adaptive class.
\end{corollary}

The benchmark emits product states conditional on the error trajectory.
The coherent advantage therefore does not require entanglement between the
emitted pulses. The relevant resource is the receiver's ability to retain
pulse-dependent quantum information. A boundary that measures its entire
input and prepares a new state can instead be simulated by retaining the
measurement outcome classically, as the next proposition establishes.

\begin{proposition}[Entanglement-breaking persistent boundaries]
\label{prop:eb-simulability}
Consider a receiver-independent temporal input and allow unrestricted
classical side information. After each intermediate round \(t<W\), let \(Q_t\)
denote the complete pulse-dependent quantum system retained for round
\(t+1\), and include any outcome produced at that boundary in the retained
classical history. Suppose that, for every incoming classical history, the
complete boundary instrument is measure-and-prepare: each of its
classical-outcome branches admits a measure-and-prepare decomposition for the
state of \(Q_t\). If no other pulse-dependent quantum system survives between
rounds, the final action statistics can be reproduced by a separate adaptive
receiver. Hence a certified advantage requires at least one intermediate
boundary whose complete, classically flagged instrument is not
entanglement breaking.
\end{proposition}

\begin{remark}
The entanglement-breaking condition applies to the complete quantum system
crossing the persistent-memory boundary. Testing a reduced memory channel
after fixing the incoming signal is insufficient: a SWAP-type update can make
that reduced map entanglement breaking while transferring the signal's quantum
information intact into the retained register.
\end{remark}

\begin{proof}
Let \(Q_0\) be a fixed, input-independent state. Suppose inductively that,
conditional on the classical history \(h\) before round \(t\), the retained
memory is a known prepared state \(\sigma_h\), uncorrelated with the
unprocessed signals. Those signals may have an arbitrary joint conditional
state. Include the boundary's classical output \(z\) in the next history.
The branchwise measure-and-prepare assumption gives
\begin{equation}
 \cE_{t,z|h}(X)
 =\int\Tr(F_{z,y|h}X)\,\sigma_{h,z,y}\,d\mu_{z|h}(y),
 \label{eq:eb-form}
\end{equation}
where \(F_{z,y|h}\succeq0\), and summing over \(z\) and integrating over
\(y\) gives \(\id\) on \(Q_{t-1}\otimes\cH_t\). Define effects on the current
signal by
\begin{equation}
 G_{z,y|h}=\Tr_{Q_{t-1}}\!\left[
 (\sigma_h^{1/2}\otimes\id)F_{z,y|h}
 (\sigma_h^{1/2}\otimes\id)\right].
 \label{eq:eb-induced-povm}
\end{equation}
They are positive and sum (with integration) to the identity on \(\cH_t\).
A separate adaptive receiver can measure this conditional POVM, retain
\((z,y)\) classically, and prepare \(\sigma_{h,z,y}\) locally. This gives
the same joint outcome distribution and conditional state of all untouched
future signals as the original instrument. It also restores the induction
hypothesis for the next round.

At the last round, let \(\{N_{a|h}\}_a\) be the final POVM on
\(Q_{W-1}\otimes\cH_W\). With the entering memory prepared in \(\sigma_h\),
the conditional effects
\begin{equation}
 B_{a|h}=\Tr_{Q_{W-1}}\!\left[
 (\sigma_h^{1/2}\otimes\id)N_{a|h}
 (\sigma_h^{1/2}\otimes\id)\right]
 \label{eq:eb-final-effect}
\end{equation}
form a POVM on the last signal and reproduce the final action statistics.
The resulting protocol measures each signal on arrival and retains only a
classical transcript. No independence assumption on the emitted temporal
state is needed.
\end{proof}

Non-entanglement-breaking memory is therefore necessary, but not sufficient,
for an advantage in this receiver model. This conclusion does not extend to
more general multitime intervention settings \citep{vieira2025eb}.

\section{Two-round separation}

\subsection{Collective and PPT programs}

To bound the entire adaptive class, we optimize over a larger measurement
class with directly checkable constraints. A feasible coherent measurement
gives an achievable error upper bound. A feasible dual witness for the larger
class gives a lower bound that applies to every adaptive receiver. If the
upper bound is smaller, no assumption of global optimality of the coherent
search is needed.

The unrestricted collective receiver minimizes the error over all POVMs on the
full two-round signal space and provides a reference for the streaming
receivers. The collective optimum is the SDP
\begin{subequations}
\begin{align}
 \underset{\{M_a\}}{\operatorname{minimize}}\quad&
 \sum_a\Tr(C_aM_a),\\
 \text{subject to}\quad&M_a\succeq0,\qquad \sum_aM_a=\id .
\end{align}
\label{eq:collective-sdp}
\end{subequations}
Its dual, with Hermitian witness \(Y\), maximizes \(\Tr Y\) subject to \(Y\preceq C_a\) for every \(a\)
\citep[Sec.~3.1.2]{watrous2018theory}.
Primal feasibility gives an achievable error; dual feasibility gives a lower
bound on the collective optimum.

Let \(\Tpt\) denote partial transpose on the first temporal factor,
\((A\otimes B)^{\mathsf T_1}=A^{\mathsf T}\otimes B\).  Because
every separate adaptive effect is separable across the two temporal factors,
it remains positive under \(\Tpt\).  Dropping separability and imposing only
this necessary PPT condition enlarges the feasible measurement class and gives
\begin{subequations}
\begin{align}
 \Perr^{\ppt}:=\underset{\{M_a\}}{\operatorname{minimize}}\quad&
 \sum_a\Tr(C_aM_a),\\
 \text{subject to}\quad&M_a\succeq0,\quad \Tpt(M_a)\succeq0,\nonumber\\
 &\sum_aM_a=\id .
\end{align}
\label{eq:ppt-sdp}
\end{subequations}
Here the dual uses a Hermitian operator \(Y\) and one positive multiplier
\(Q_a\) for each action.

A convenient dual is
\begin{subequations}
\begin{align}
 \underset{Y=Y^\dagger,\{Q_a\}}{\operatorname{maximize}}\quad&\Tr Y,\\
 \text{subject to}\quad&Q_a\succeq0,\nonumber\\
 &C_a-Y-\Tpt(Q_a)\succeq0\quad\forall a .
\end{align}
\label{eq:ppt-dual}
\end{subequations}
This is the loss-minimization form of the standard PPT discrimination
primal--dual pair \citep[Sec.~3.1]{cosentino2013ppt}; see also
\citep[Sec.~1.2.3]{watrous2018theory} for SDP duality.

\begin{proposition}[Adaptive containment]
\label{prop:containment}
Every two-round separate adaptive receiver induces a POVM feasible for
\cref{eq:ppt-sdp}.  Consequently,
\begin{equation}
 \Perr^{\ppt}\le \Padstar(C).
 \label{eq:ordering}
\end{equation}
\end{proposition}

\begin{proof}
Each coarse-grained effect is an integral of positive product operators, as in
\cref{eq:adaptive}.  It is therefore separable.  Partial transpose maps
\(A\otimes B\) to \(A^{\mathsf T}\otimes B\succeq0\); integration, convex
mixing, local ancillas, continuous records, and randomization preserve this
property.  The completeness relation is unchanged.
\end{proof}

A coherent receiver whose risk lies below a certified lower bound on
\(\Perr^{\ppt}\) therefore outperforms the entire separate adaptive class.

Floating-point optimization is used to find candidate receivers and dual
witnesses, but it is not used directly for the theorem. The numerical receiver
and dual candidates are embedded as exact rational data and checked against
cost operators reconstructed directly from the analytic benchmark using
256-bit Arb/Acb ball arithmetic. Outward
rounding yields a risk upper bound for each explicit receiver and a dual
objective lower bound for the PPT relaxation. The procedure is described in
Appendix~\ref{app:certificate}.

\subsection{Separation criterion and numerical values}

\begin{theorem}[Certified two-round separation]
\label{thm:main}
For the two-round cost operators \(C\), the certified \(K=7\) streaming
receiver and PPT-dual witness satisfy
\begin{gather}
 r(M_{K=7};C)\le U_7:=0.242099929928125,\nonumber\\
 L_{\rm PPT}:=0.251431736761034\le\Padstar(C).
 \label{eq:w2-safe-pair}
\end{gather}
Consequently,
\[
 \Padstar(C)-\Pcohstar(C)
 \ge L_{\rm PPT}-U_7
 \ge0.009331806832908.
\]
\end{theorem}

The upper bound \(U_7\) concerns one explicitly feasible receiver, whereas
\(L_{\rm PPT}\) lower-bounds the error of every adaptive receiver. Their
strict ordering certifies the class separation. They are bounds rather than
claims that both class optima are known exactly.

The theorem follows from weak duality and feasibility of the explicit
receiver. The lower bound applies to every separate adaptive receiver;
tightness of the PPT relaxation is unnecessary. The same calculation encloses
the collective optimum as
\begin{equation}
 0.242099896126073\le \Pcollstar(C)
 \le0.242099929928125.
 \label{eq:collective-enclosure}
\end{equation}
The \(K=7\) receiver is therefore within
\(3.3802052\times10^{-8}\) of the collective optimum. The enclosure does not
establish equality with the collective optimum.

\begin{table}[t]
\centering
\small
\setlength{\tabcolsep}{4pt}
\caption{\textbf{Certified two-round benchmark.} Inequality directions follow
from outward-rounded receiver and dual verification.}
\label{tab:headline}
\resizebox{\columnwidth}{!}{%
\begin{tabular}{@{}lr@{}}
\toprule
Quantity & Recovery-decision error\\
\midrule
Collective optimum & \([0.242099896126073,\;0.242099929928125]\)\\
Certified \(K=7\) risk upper bound & \(\le0.242099929928125\)\\
Certified PPT lower bound on adaptive error & \(\ge0.251431736761034\)\\
Achievable adaptive risk upper bound & \(\le0.256727930428793\)\\
\midrule
Certified gap lower bound & \(\ge0.009331806832908\)\\
\bottomrule
\end{tabular}
}
\end{table}

\begin{figure}[t]
\centering
\safegraphic{0.96\columnwidth}{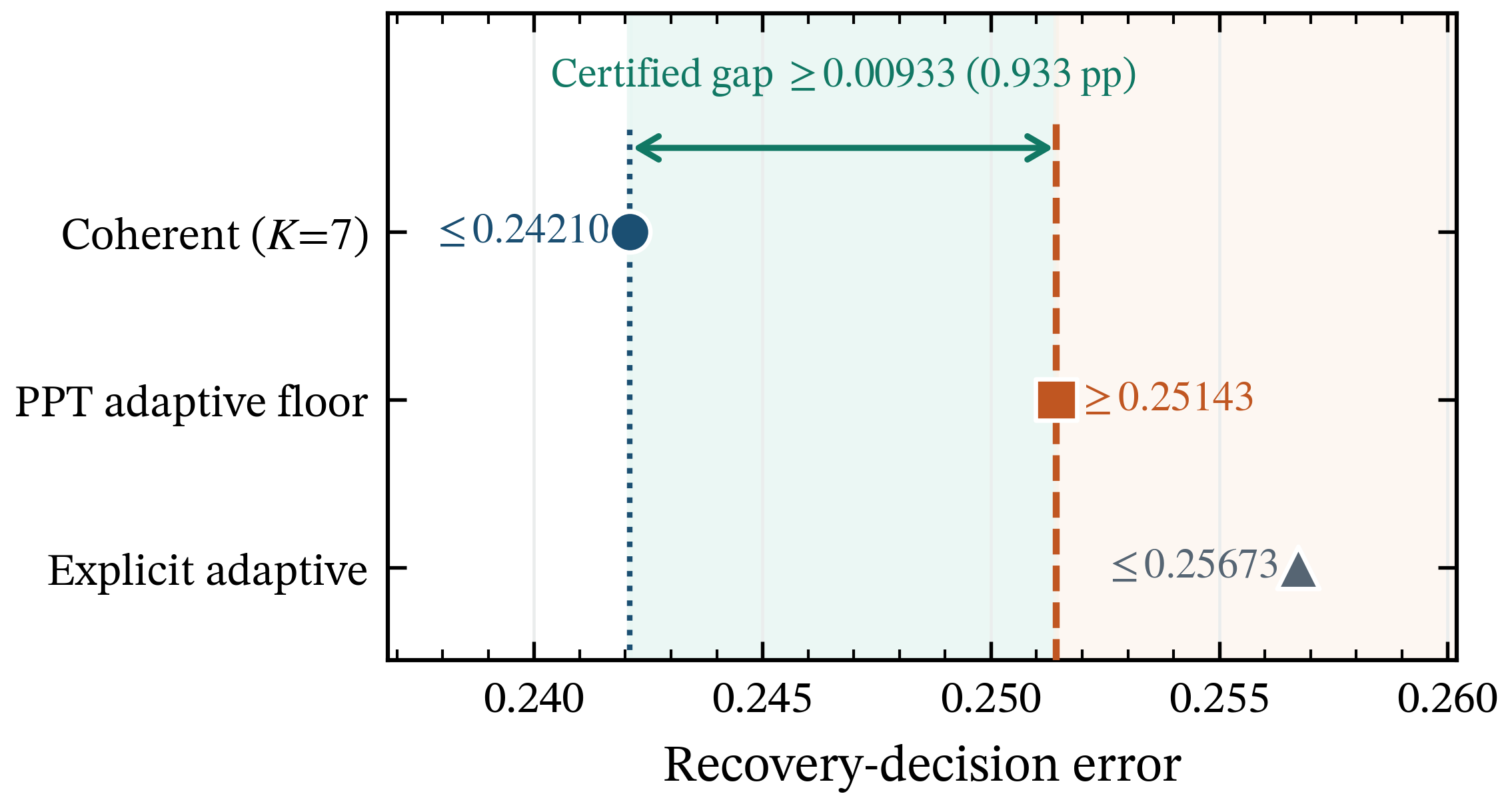}
\caption{\textbf{Certified two-round comparison.} The horizontal axis is
recovery-decision error. The coherent-risk upper
bound lies below the certified PPT lower bound on the separate-adaptive
optimum. The explicit adaptive receiver is shown as an achievable reference.}
\label{fig:headline}
\end{figure}

\begin{proposition}[Stability of a certified separation]
\label{prop:stability}
Let \(C'=\{C'_a\}\) be a perturbed decision problem on the same physical
Hilbert space and with the same receiver classes, where
\(C'_a=C_a+\Delta C_a\).  Suppose \(L_{\adapt}\le\Padstar(C)\) is a certified
lower bound and a fixed feasible coherent receiver \(M^{\rm coh}\) has risk
\(r(M^{\rm coh};C)\).  Define
\begin{equation}
 \epsilon_*(C',C):=\inf_{X=X^\dagger}
 \sum_a\lVert\Delta C_a-X\rVert_1,
 \label{eq:epsilon-star}
\end{equation}
where the common Hermitian shift \(X\) is independent of \(a\).  If
\(\delta:=L_{\adapt}-r(M^{\rm coh};C)>0\), then the perturbed class optima obey
\begin{equation}
 \Padstar(C')-\Pcohstar(C')
 \ge \delta-2\epsilon_*(C',C).
 \label{eq:stability}
\end{equation}
\end{proposition}

\begin{proof}
For any POVM \(\{M_a\}\), \(0\preceq M_a\preceq\id\), and hence, for every
Hermitian \(X\),
\begin{equation}
 \left|\sum_a\Tr[(\Delta C_a-X)M_a]\right|
 \le \sum_a\lVert\Delta C_a-X\rVert_1=:\epsilon_X .
 \label{eq:stability-holder}
\end{equation}
POVM completeness makes the common shift contribute
\(\sum_a\Tr(XM_a)=\Tr X\).  Therefore
\(\Padstar(C')\ge L_{\adapt}+\Tr X-\epsilon_X\).
Evaluating the same explicit coherent receiver on \(C'\) gives
\(r(M^{\rm coh};C')\le r(M^{\rm coh};C)+\Tr X+\epsilon_X\), while
\(\Pcohstar(C')\le r(M^{\rm coh};C')\).  Subtraction and optimization over
\(X\) prove
\cref{eq:stability}.
\end{proof}

For the certified two-round gap, Proposition~\ref{prop:stability} guarantees
separation whenever
\begin{equation}
 \epsilon_*<0.004665903416454 .
 \label{eq:stability-radius}
\end{equation}
The bound is sufficient and need not be tight.

\subsection{Certified parameter neighborhood}

Interval verification gives the same strict two-round threshold throughout
\begin{equation}
 p\in[0.29999,0.30001],\qquad
 \alpha\in[0.59999,0.60001],
 \label{eq:k4-parameter-box}
\end{equation}
with a whole-box separation lower bound of \(0.001627522285899\). Thus
\(K_{\rm sep}=4\) throughout this inner rectangle in the strict model. The
verification includes both exclusion of \(K=1,2,3\), with a uniform margin of
at least \(0.034644013557275\), and a feasible \(K=4\) receiver throughout the
rectangle. A separate \(K=7\) certificate covers
\begin{equation}
 p\in[0.2999,0.3001],\qquad
 \alpha\in[0.5999,0.6001],
 \label{eq:k7-parameter-box}
\end{equation}
with gap at least \(0.001932932526545\) and uniform \(K\le3\) exclusion margin
at least \(0.023874515270248\); there the established memory statement
is \(4\le K_{\rm sep}\le7\). All values of \(\alpha\) are represented in a
common analytic signal frame; details are given in
Appendix~\ref{app:benchmark}. These rectangles give local
parameter-robustness certificates around the benchmark point.

\section{Temporal quantum-memory complexity}

\subsection{Streaming optimization}

A collective POVM does not specify the sequential channels needed for a
streaming implementation.  We therefore
optimize the channel sequence in \cref{eq:streammaps} using Choi-matrix
representations, enforcing complete positivity and trace preservation for each
update and positivity and completeness for the final POVM.  With all other
blocks fixed, one update channel or the final measurement enters the objective
linearly and can be optimized by an SDP. Each completed sweep produces a
candidate receiver, which is subsequently checked and, when
necessary, repaired as described in Appendix~\ref{app:choi}.
The joint problem is nevertheless nonconvex, so the reported values are the
best feasible values found rather than certified global optima.
Appendix~\ref{app:choi} states the Choi convention, block structure, and
feasibility checks used in this calculation.

A Choi matrix represents a channel as a positive matrix subject to a linear
trace-preservation condition. This is why each individual update can be
optimized by an SDP when the others are fixed. Its convention is given in
Appendix~\ref{app:choi}.

For the strict class in Definition~\ref{def:strict-streaming}, define
\begin{align}
 P_{K,\rm str}^{\star}(C)
 &:=\min_{M\in\cM_{\rm str}^{(K)}}r(M;C),
 \label{eq:pk-strict}\\
 K_{\rm sep}(C)
 &:=\min\{K:P_{K,\rm str}^{\star}(C)<\Padstar(C)\},
 \label{eq:k-sep}\\
 K_{\rm coll}(C)
 &:=\min\{K:P_{K,\rm str}^{\star}(C)=\Pcollstar(C)\}.
 \label{eq:k-coll}
\end{align}
The minimum of an empty set is \(\infty\). Thus \(K_{\rm sep}\) is the
smallest dimension whose optimal strict receiver outperforms the adaptive
optimum; the certificates determine it by upper and lower bounds. It is not
defined by the outcome of a particular local search. These definitions apply to the
strict counted-register convention and do not include architectures with an
additional persistent classical record.

\begin{proposition}[Attainment and monotonicity]
\label{prop:memory-monotonicity}
For finite signal and action spaces, the minimum in
Eq.~\eqref{eq:pk-strict} is attained and
\begin{equation}
 P_{K+1,\rm str}^{\star}(C)\le P_{K,\rm str}^{\star}(C).
 \label{eq:memory-monotonicity}
\end{equation}
\end{proposition}

\begin{theorem}[Constructive full-storage bound]
\label{thm:full-storage}
For \(W\) signals of common dimension \(d_{\mathrm{sig}}\) and \(A\) final actions, every
\(A\)-outcome collective POVM can be realized by a strict streaming receiver
with
\begin{equation}
 K=\max\{d_{\mathrm{sig}}^{W-1},A\}.
 \label{eq:full-storage-bound}
\end{equation}
\end{theorem}

The operator-Schmidt rank \(\operatorname{OSR}_t(M_a)\) is the smallest
number of product operators \(L_\mu\otimes R_\mu\) needed to express the
effect \(M_a\) across the cut between the first \(t\) signals and the rest.

\begin{theorem}[Finite-memory operator-Schmidt-rank bound]
\label{thm:osr}
Every terminal effect \(M_a\) induced by a strict memory-\(K\) receiver obeys
\begin{equation}
 \operatorname{OSR}_t(M_a)\le K^2
 \label{eq:osr-bound}
\end{equation}
across every chronological prefix-suffix cut
\(\cH_{1:t}|\cH_{t+1:W}\). For a fixed receiver, the action effects share the
same channel-generated left slice space at each chronological cut.
\end{theorem}

The \(K^2\)-dimensional operator space of a memory limits the structure that
can link earlier and later signals in each decision effect. This rank bound
describes the receiver's measurement, not entanglement of the emitted input.

The proofs are given in Appendix~\ref{app:memory-complexity}. The
operator-Schmidt statement is the standard matrix-product-operator bond
constraint associated with a finite operator space
\citep{verstraete2004mpdo,schollwock2011dmrg}; its use here is to
formulate memory requirements for the recovery-weighted decision problem.

\subsection{Global lower bounds for fixed memory dimension}
\label{sec:fixedK}

The minimality statements below require lower bounds valid for \emph{every}
strict memory-\(K\) receiver, not merely the best value found by a local
search. Two such bounds are used. Both are stated with the success operators
\begin{equation}
 S_a:=\sum_{x\in\cX}\pi_x[1-\ell(a,x)]\rho_x ,
 \label{eq:success-op}
\end{equation}
which satisfy \(\sum_a\Tr(S_aM_a)=1-r(M;C)\) for the binary loss.

The success operator \(S_a\) weights each trajectory state by its reward for
action \(a\). In the relaxation below, \(F_a\) is an effective input effect,
\(t_a\) bounds its operator norm, and the sum of these bounds is limited by
the terminal register dimension.

\begin{lemma}[Terminal norm budget]
\label{lem:norm-budget}
Let \(\{F_a\}\) be the effective POVM on the full temporal input induced by a
strict memory-\(K\) receiver. Then
\begin{equation}
 \sum_a\lVert F_a\rVert_\infty\le K .
 \label{eq:norm-budget}
\end{equation}
\end{lemma}

\begin{proof}
Write \(F_a=\Phi^\dagger(N_a)\), where \(\Phi\) is the composed channel from
the temporal input to the terminal \(K\)-dimensional register and \(N_a\) is
the terminal POVM element. The adjoint \(\Phi^\dagger\) is positive and
unital, so \(\lVert F_a\rVert_\infty\le\lVert N_a\rVert_\infty\). Since
\(N_a\succeq0\), \(\lVert N_a\rVert_\infty\le\Tr N_a\), and
\(\sum_a\Tr N_a=\Tr\id_K=K\).
\end{proof}

Lemma~\ref{lem:norm-budget} gives the convex outer relaxation
\begin{subequations}
\begin{align}
 \underset{\{F_a\},\{t_a\}}{\operatorname{maximize}}\quad&
   \sum_a\Tr(S_aF_a),\\
 \text{subject to}\quad&F_a\succeq0,\qquad\sum_aF_a=\id,\\
 &F_a\preceq t_a\id,\qquad t_a\ge0,\\
 &\sum_a t_a\le K .
\end{align}
\label{eq:norm-budget-primal}
\end{subequations}
Every strict memory-\(K\) receiver is feasible for
\cref{eq:norm-budget-primal}, so its optimum upper-bounds the achievable
success probability. A convenient dual is
\begin{subequations}
\begin{align}
 \underset{Y=Y^\dagger,\{Z_a\},\mu}{\operatorname{minimize}}\quad&\Tr Y+K\mu,\\
 \text{subject to}\quad&Z_a\succeq0,\\
 &Y+Z_a-S_a\succeq0,\\
 &\Tr Z_a\le\mu,\qquad\mu\ge0 .
\end{align}
\label{eq:norm-budget-dual}
\end{subequations}
The dual follows by assigning a Hermitian multiplier \(Y\) to POVM
completeness, positive multipliers \(Z_a\) to \(F_a\preceq t_a\id\),
and a nonnegative multiplier \(\mu\) to the terminal norm budget,
using standard SDP duality \citep[Sec.~1.2.3]{watrous2018theory}.
Any certified feasible dual value \(D_K\) therefore gives the global bound
\begin{equation}
 P_{K,\rm str}^{\star}(C)\ge 1-D_K .
 \label{eq:global-floor}
\end{equation}
For the two-round benchmark, interval verification of feasible duals gives
\begin{equation}
\begin{aligned}
 P_{1,\rm str}^{\star}&=0.48,\\
 P_{2,\rm str}^{\star}&\ge0.360464844234613,\\
 P_{3,\rm str}^{\star}&\ge0.292565480672402 .
\end{aligned}
 \label{eq:w2-global-floors}
\end{equation}

In the dual, \(Y\) is Hermitian, \(Z_a\) are positive multipliers, and
\(\mu\) is a nonnegative scalar. A feasible value \(D_K\) upper-bounds
success, so \(1-D_K\) lower-bounds the risk.

A second, purely combinatorial bound is useful at greater depth.

\begin{lemma}[Terminal-dimension prior bound]
\label{lem:dimension-prior}
For binary loss \(\ell(a,x)\in\{0,1\}\), let
\(\omega_a=\sum_x\pi_x[1-\ell(a,x)]\), and order these weights as
\(\omega_{(1)}\ge\omega_{(2)}\ge\cdots\), padding with zeros if
\(K>\lvert\cA\rvert\). Then
\begin{equation}
 P_{K,\rm str}^{\star}(C)\ge
 1-\sum_{j=1}^{K}\omega_{(j)} .
 \label{eq:dimension-prior}
\end{equation}
\end{lemma}

The proof is given in Appendix~\ref{app:memory-complexity}.

\subsection{Memory-dimension frontier}

\begin{figure}[t]
\centering
\safegraphic{\columnwidth}{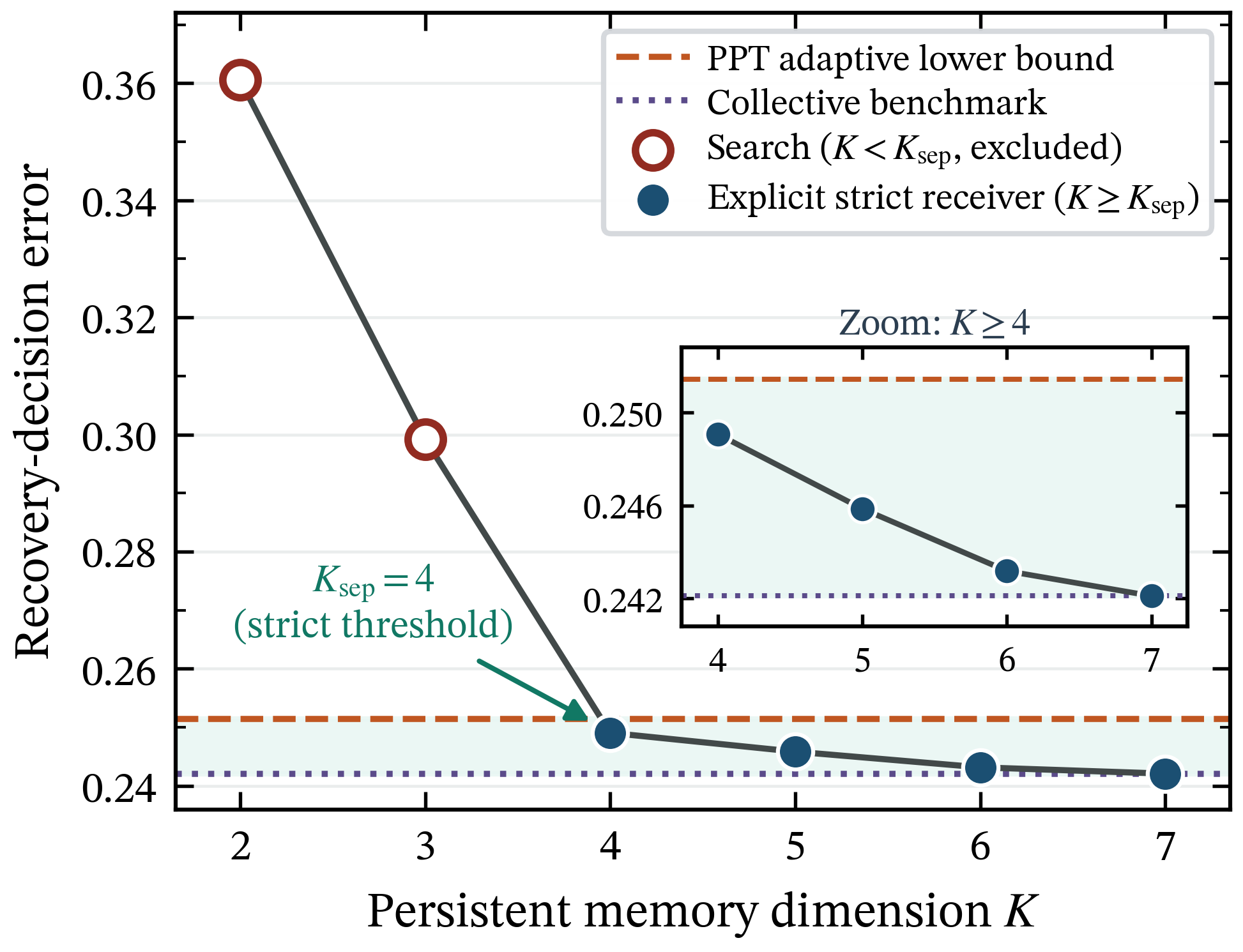}
\caption{\textbf{Two-round memory-performance search.} The horizontal axis
is memory dimension \(K\), and the vertical axis is recovery-decision error.
Markers show the best
values found at each tested dimension. The exact strict-model threshold
\(K_{\rm sep}=4\) follows from the global exclusion certificates in
Table~\ref{tab:memory}, not from the plotted \(K=2,3\) points. The plotted
collective line is the numerical estimate used in the search;
Eq.~\eqref{eq:collective-enclosure} gives the certified enclosure.}
\label{fig:memory}
\end{figure}

Figure~\ref{fig:memory} shows the best values obtained in the fixed-memory
search, while Table~\ref{tab:memory} gives the separate certified bounds used
to establish the strict memory threshold. The plotted \(K=2\) and \(K=3\)
values are feasible search points; their exclusion instead follows from
\cref{eq:w2-global-floors}. The \(K=4\) and \(K=7\) entries are explicit
receivers whose risks are certified by outward rounding.

\begin{table}[t]
\centering
\caption{\textbf{Certified bounds used for the two-round strict-memory
threshold.} The adaptive entry is an achievable risk upper bound.}
\label{tab:memory}
\resizebox{\columnwidth}{!}{%
\begin{tabular}{@{}cll@{}}
\toprule
Object & Certified relation & Role\\
\midrule
\(P_{1,\rm str}^{\star}\) & \(=0.48\) & global value\\
\(P_{2,\rm str}^{\star}\) & \(\ge0.360464844234613\) & global lower bound\\
\(P_{3,\rm str}^{\star}\) & \(\ge0.292565480672402\) & global lower bound\\
\(\Padstar(C)\) & \(\le U_{\rm ad}:=0.256727930428793\) & adaptive strategy\\
\(r(M_{K=4};C)\) & \(\le U_4:=0.249065686566581\) & explicit strict receiver\\
\(\Padstar(C)\) & \(\ge L_{\rm ad}:=0.251431736761034\) & PPT lower bound\\
\bottomrule
\end{tabular}
}
\end{table}

For \(K=1,2,3\), global lower bounds exceed the error of an available
adaptive strategy, excluding an advantage at these dimensions. A certified
\(K=4\) receiver instead has error below the lower bound for every adaptive
receiver. Together these comparisons establish necessity and sufficiency.
Four dimensions could be represented by two ideal qubits; the threshold
counts the entire persistent and terminal register in the strict model.

\begin{theorem}[Two-round strict-memory threshold]
\label{thm:w2-memory-threshold}
For the certified two-round benchmark and the strict terminal-register model,
\begin{equation}
 K_{\rm sep}=4,
 \qquad 4\le K_{\rm coll}\le7 .
 \label{eq:w2-memory-threshold}
\end{equation}
\end{theorem}

\begin{proof}
The first three rows of Table~\ref{tab:memory} give
\(P_{K,\rm str}^{\star}>U_{\rm ad}\) for \(K=1,2,3\). Since the displayed
adaptive strategy is feasible, \(\Padstar\le U_{\rm ad}\); therefore none of
these dimensions can beat the adaptive optimum. Conversely, the certified
\(K=4\) receiver obeys
\begin{equation}
 L_{\rm ad}-U_4\ge0.002366050194452>0,
 \label{eq:k4-safe-gap}
\end{equation}
and \(L_{\rm ad}\le\Padstar\), so \(K=4\) suffices. This proves
\(K_{\rm sep}=4\), and the same exclusion gives \(K_{\rm coll}\ge4\).
Recovery action \(7\) is never correct for the two-round benchmark, so its
success operator satisfies \(S_7=0\). Any weight assigned to outcome \(7\) can
be reassigned to another outcome without decreasing the success probability,
since all success operators are positive semidefinite. Hence an optimal
collective POVM may be chosen with at most seven nonzero decision outcomes.
Applying
Theorem~\ref{thm:full-storage} with \(d_{\mathrm{sig}}=4\) and \(A=7\) gives
\(K_{\rm coll}\le7\).
\end{proof}

The threshold refers to the strict counted-register convention: the terminal
register is included in \(K\), and no additional persistent classical history
is available outside that register. By
Proposition~\ref{prop:eb-simulability}, the certified \(K=4\) receiver must
contain a non-entanglement-breaking persistent boundary. This is a statement
about the channel, not about entanglement of the benchmark state.

\section{Multi-round finite-memory results}

For \(W>2\), temporal separability again permits PPT constraints across chosen
temporal bipartitions. The \(W=3\) certificate imposes all three one-versus-two
cuts simultaneously (only \(1|23\) and \(12|3\) are chronological),
\begin{equation}
 1|23,\qquad2|13,\qquad3|12,
 \label{eq:w3-cuts}
\end{equation}
and gives \(L_{3,\rm PPT}\ge0.299647463773192\). Only the balanced cut
\(12|34\) is imposed in the \(W=4\) calculation, which gives
\(L_{4,\rm PPT}\ge0.319542490402679\).

The full-storage construction of Theorem~\ref{thm:full-storage} gives
\(K=16\) at \(W=3\) and \(K=64\) at \(W=4\) for \(d_{\mathrm{sig}}=4\) and \(A=8\).
Certified separation is achieved with \(K=8\) and \(K=36\),
respectively. Table~\ref{tab:compressed-depth} collects the outward-safe
bounds.

\begin{table*}[t]
\centering
\caption{\textbf{Certified streaming constructions at greater depth.} The
\(W=3\) comparator uses all cuts in Eq.~\eqref{eq:w3-cuts}; the \(W=4\)
comparator uses only \(12|34\).}
\label{tab:compressed-depth}
\begin{tabular}{@{}ccccc@{}}
\toprule
Depth & Memory & Certified risk upper bound & PPT lower bound & Gap lower bound\\
\midrule
\(W=3\) & \(K=8\)  & \(0.294297344448216\) & \(0.299647463773192\) & \(0.005350119324975\)\\
\(W=3\) & \(K=16\) & \(0.284502699489212\) & \(0.299647463773192\) & \(0.015144764283980\)\\
\(W=4\) & \(K=36\) & \(0.319370057047514\) & \(0.319542490402679\) & \(0.000172433355164\)\\
\(W=4\) & \(K=64\) & \(0.317529609881821\) & \(0.319542490402679\) & \(0.002012880520857\)\\
\bottomrule
\end{tabular}
\end{table*}

The \(K=16\) and \(K=64\) entries are explicit strict streaming
realizations. The \(K=8\) and \(K=36\) entries are lower-memory streaming
receivers obtained directly from the restricted prefix-memory construction
described in Appendix~\ref{app:multi-round}.

\begin{table}[t]
\centering
\caption{\textbf{Current strict-memory bounds.} Only the two-round separation
threshold is exact. The prefix column is \(d_{\mathrm{sig}}^{W-1}\), before accounting for
the terminal action register in Eq.~\eqref{eq:full-storage-bound}.}
\label{tab:memory-depth}
\begin{tabular}{@{}cccc@{}}
\toprule
\(W\) & \(K_{\rm sep}\) & \(K_{\rm coll}\) & Prefix dimension\\
\midrule
2 & \(4\) & \([4,7]\) & 4\\
3 & \([3,8]\) & \([4,16]\) & 16\\
4 & \([3,36]\) & \([4,64]\) & 64\\
\bottomrule
\end{tabular}
\end{table}

The lower endpoints at \(W=3\) and \(W=4\) follow from
Lemma~\ref{lem:dimension-prior}.  Certified feasible product receivers give
the adaptive-class upper bounds
\begin{align*}
 P_{\mathrm{ad}}^{\star}(W=3)&\le0.405392108675551,\\
 P_{\mathrm{ad}}^{\star}(W=4)&\le0.444280960818635 .
\end{align*}
The dimension-prior floor excludes \(K=1,2\) against these feasible
comparators, giving \(K_{\rm sep}\ge3\) at both depths.  The corresponding
\(K=3\) dimension-prior lower bounds lie above the certified feasible
full-storage risk upper bounds \(0.284502699489212\) at \(W=3\) and
\(0.317529609881821\) at \(W=4\), establishing \(K_{\rm coll}\ge4\).

The reduction from \(64\) to \(36\) is in Hilbert-space dimension.
Both need six qubits in a binary embedding:
\(\lceil\log_2 36\rceil=\lceil\log_2 64\rceil=6\).
A smaller dimension alone does not imply fewer physical qubits or a simpler
control circuit.

At \(W=3\) and \(W=4\), full-prefix storage is therefore not necessary for a
certified separation. The tested depths do not determine asymptotic memory
scaling. The certified margins are also not monotonic in \(W\): the stronger
three-cut \(W=3\) relaxation produces a larger full-storage margin than the
two-round calculation, while the present \(W=4\) bound uses only one balanced
cut. The margin therefore depends on both the recovery problem and the
tightness of the chosen outer relaxation. Appendix~\ref{app:multi-round}
describes the deeper streaming constructions and certificates; the
observation-only control remains separate in
Appendix~\ref{app:observation-control}.

\section{Robustness and implementation error}
\label{sec:robustness}

\subsection{Dephasing of the counted persistent register}

We first apply noise to the register of an explicit streaming receiver while
holding its channels and terminal measurement fixed. This tests whether its
advantage survives imperfections. Section~\ref{sec:hardware} instead acts on
the original pulse registers and reoptimizes a collective measurement.

For the strict streaming receivers, whole-register dephasing is applied after
each intermediate update. The update channels and terminal POVM are otherwise
kept fixed. We use \(q\) for dephasing strength and \(\eta=1-q\) for coherence
retention:
\begin{equation}
 \mathcal D_q(\rho)=(1-q)\rho
 +q\sum_{m=0}^{K-1}\ket m\!\bra m\rho\ket m\!\bra m .
 \label{eq:persistent-dephasing}
\end{equation}
This differs from reoptimizing a terminal measurement for each noise strength.
At \(W=2\) there is a single intermediate memory boundary, so the certified
curve is a test of inter-round memory alone and the risk is affine in \(q\).

Interval verification brackets the crossing points near
\begin{equation}
\begin{aligned}
 q_\star^{(K=4)}&\approx0.0169123753455,\\
 q_\star^{(K=7)}&\approx0.0519338246576 .
\end{aligned}
 \label{eq:persistent-crossings}
\end{equation}
The available bracket information is listed in Appendix~\ref{app:robustness};
exact two-round endpoints remain to be transcribed.
Thus the fixed \(K=4\) and \(K=7\) receivers retain a certified advantage up
to approximately \(1.69\%\) and \(5.19\%\) dephasing strength, respectively.
These dephasing strengths are defined in the
numerical memory basis and are not calibrated device error rates.
Analogous fixed-receiver certificates are obtained for the deeper
\(W=3,K=8\) and \(W=4,K=36\) constructions; their exact crossing brackets are
listed in Appendix~\ref{app:robustness}.

\subsection{Channel-distance error budget}

\begin{theorem}[Implementation-error accumulation]
\label{thm:diamond-budget}
Let the ideal strict receiver be a composition of CPTP factors, including a
terminal measurement channel, and let factor \(j\) be replaced by an
implementation at unhalved diamond distance
\(\delta_j=\lVert\widetilde{\cE}_j-\cE_j\rVert_\diamond\). The diamond norm
maximizes the output trace norm over joint inputs with an arbitrary reference
system and input trace norm one. No factor of one half is included in
\(\delta_j\). For any loss in
\([0,1]\),
\begin{equation}
 \left|r_{\rm impl}-r_{\rm ideal}\right|
 \le\min\left\{1,\frac12\sum_j\delta_j\right\}.
 \label{eq:diamond-risk-bound}
\end{equation}
\end{theorem}

The proof is a telescoping hybrid argument followed by contractivity of trace
distance and is given in Appendix~\ref{app:robustness}. Combining
Eq.~\eqref{eq:diamond-risk-bound} with the certified W2 margins gives the
sufficient budgets
\begin{equation}
\begin{aligned}
 \sum_j\delta_j&<0.004732100388904 &&(K=4),\\
 \sum_j\delta_j&<0.018663613665816 &&(K=7).
\end{aligned}
 \label{eq:w2-diamond-budgets}
\end{equation}
These are dimensionless sufficient channel-error budgets, not calibrated
hardware tolerances.

\section{Stored-pulse hardware model}
\label{sec:hardware}

We also consider a stored-pulse model in which individual pulse registers
undergo phase decay before an optimized collective readout. This model acts on
the stored pulse registers rather than on the learned \(K\)-dimensional
streaming memory. At each trial value of \(T_2\), the
final joint POVM is reoptimized while the stored pulses undergo
pulse-dependent phase retention.

Here \(T_2\) is the coherence-decay time, \(T_1\) the energy-relaxation time,
and \(T_\phi\) the pure-dephasing time. The interval
\(\tau_{\rm round}\) separates pulses, and \(\tau_{\rm proc}\) is the final
processing delay.
For pulse \(t\in\{1,\ldots,W\}\), the modeled storage time and retention are
\begin{equation}
 t_t=\tau_{\rm proc}+(W-t)\tau_{\rm round},
 \qquad \lambda_t=e^{-t_t/T_2}.
 \label{eq:t2}
\end{equation}
The earliest pulse has the longest modeled storage time,
\begin{equation}
 t_1=\tau_{\rm proc}+(W-1)\tau_{\rm round}.
 \label{eq:t1max}
\end{equation}
Amplitude loss must be simulated separately, for example with
\begin{equation}
 \eta_{1,t}=e^{-t_t/T_1}.
 \label{eq:t1}
\end{equation}
Although
\begin{equation}
 \frac{1}{T_2}=\frac{1}{2T_1}+\frac{1}{T_\phi}
 \label{eq:t1t2}
\end{equation}
relates the decay constants in a Markovian model, it does not replace separate
amplitude-damping and pure-dephasing channels.

For the slow schedule, the timing anchors are
\begin{equation}
\begin{aligned}
 \tau_{\mathrm{round}}&=5.82\,\mu\mathrm{s},\\
 \tau_{\mathrm{proc}}&=2.64\,\mu\mathrm{s}.
\end{aligned}
 \label{eq:storage-schedule}
\end{equation}
The timing parameters are modeling inputs drawn from the stated schedule and
a literature anchor \citep{riste2012feedback}; they are not measurements of an
integrated implementation. The composition of \(\tau_{\mathrm{round}}\) is
given in Appendix~\ref{app:hardware-sensitivity}.
The corresponding longest storage times \(t_1\) are \(8.46\), \(14.28\), and
\(20.10\,\mu\mathrm{s}\) for \(W=2,3,4\). At a numerical crossing
\(T_{2,\star}\) we report the retention of the earliest pulse,
\begin{equation}
 \lambda_\star:=\exp(-t_1/T_{2,\star}).
 \label{eq:lambda-star}
\end{equation}
The resulting crossings, shown in Fig.~\ref{fig:hardware-t2} of
Appendix~\ref{app:hardware-sensitivity}, are
\(T_{2,\star}=259.942\,\mu\mathrm{s}\), \(767.970\,\mu\mathrm{s}\), and
\(2656.778\,\mu\mathrm{s}\) for \(W=2,3,4\).

These crossings are numerical outputs of the stored-pulse model and are not
part of the interval-certified receiver bounds.  The \(W=2\) crossing lies
within reported component-scale coherence times, whereas the \(W=3\) and
\(W=4\) calculations require progressively longer storage.  The comparison is
made at the component level; reported component values and the timing
assumptions behind them are collected in
Appendix~\ref{app:hardware-sensitivity}.

\section{Discussion}

The certified coherent-risk upper bound lies below the certified PPT lower
bound on the entire separate-adaptive class.  That class already permits
continuous outcomes, unrestricted classical memory, and unrestricted classical
computation.  The operational distinction is therefore whether pulse-dependent
quantum information may persist between rounds.

\Cref{prop:stability}
shows that the positive gap persists under
sufficiently small perturbations of the cost operators.
\Cref{prop:eb-simulability} shows
that a receiver whose complete classically flagged boundary instruments
are all measure-and-prepare can be simulated by a separate adaptive receiver.
The certified \(K=4\) separation therefore implies that at least one
persistent boundary of that receiver is not measure-and-prepare. This implication is specific to the receiver
architecture defined here; entanglement-breaking dynamics can behave
differently under more general multitime interventions
\citep{vieira2025eb}.

For the strict two-round architecture, the minimum terminal memory dimension
required to outperform the separate adaptive class is exactly four. This
threshold includes the terminal register and assumes that no additional
persistent classical history is available outside the counted register. The
fixed-receiver dephasing study acts on this counted register,
while the diamond-norm theorem bounds the effect of imperfect channel
implementations without choosing a dephasing basis.

At greater depth the exact threshold is not known, but certified separation is
already obtained with \(K=8\) at \(W=3\) and \(K=36\) at \(W=4\), compared
with \(K=16\) and \(K=64\) for the full-prefix construction. At \(W=3\), the
simultaneous three-cut relaxation gives a full-storage certified margin of
\(0.015144764283980\), while the \(W=4\) calculation presently uses only one
balanced cut. The tested depths therefore do not establish a monotonic trend
in the true coherent-adaptive gap or an asymptotic memory scaling law.

The architecture also changes where latency enters the recovery pipeline. A
separate adaptive receiver converts each arriving signal into a classical record before any
pulse-dependent information from that signal can participate in later-round
processing, whereas a coherent receiver may carry such information forward in a
quantum memory. This can replace some inter-round measurement and feed-forward
requirements with coherent updates and storage. The present results do not
establish lower end-to-end latency: coherent capture, memory operations,
storage, and final readout introduce their own timing requirements, and, in the
separate stored-pulse model considered here, the required coherence time
increases substantially with depth. A quantitative comparison of these resources would
require a hardware-specific timing model and is left for future work.

Several limitations remain. The pulse ensemble, extraction circuit,
QPSK encoding, prior, and recovery loss are fixed. PPT is an outer relaxation
and may become loose at larger temporal depth; Appendix~\ref{app:sep-example}
gives an analytic two-round task with a strict coherent advantage whose
optimal measurement is nevertheless separable.  The optimized channels have
not been compiled to a low-depth device Hamiltonian.  The QRC calculation is
exploratory, and the stored-pulse hardware
calculation is distinct from the learned streaming-memory architecture. The
most immediate extensions are a stronger multi-cut \(W=4\) relaxation,
device-level compilation of the coherent updates, and tests with independently
calibrated amplitude and phase noise.
Appendix~\ref{app:flagged} states a conditional flagged-sector inheritance
criterion, but its hypotheses have not been verified for a natural
surface-code family.

\section{Conclusion}

The point at which syndrome information becomes classical is itself a
receiver-design decision. For the fixed recovery problem studied here, an
explicit coherent streaming receiver has certified recovery risk below every
separate adaptive receiver. In the strict counted-register model, the exact
two-round separation threshold is \(K_{\rm sep}=4\), and the induced temporal
effects have operator-Schmidt rank at most \(K^2\) across every chronological
cut. At greater depth,
certified \(K=8\) and \(K=36\) receivers show that full-prefix storage is not
required for separation at \(W=3\) and \(W=4\).

Dephasing of the actual persistent register gives a direct robustness test for
the explicit receivers, while the channel-distance bound gives a general
sufficient implementation-error condition. Both results are defined within the
receiver model used here. The distinction from the separate adaptive class is
the ability to retain pulse-dependent quantum information across rounds.
Exact W3 and W4 thresholds,
a stronger multi-cut W4 bound, and device-level compilation of the certified
updates remain open.

\appendix

\section{Proof of temporal separability}
\label{app:temporal-proof}

\begin{proof}[Proof of Theorem~\ref{thm:temporal-separability}]
Condition on the receiver's private randomness. Fresh local ancillas can
be absorbed into the effective POVM on the current signal. At each node of
the adaptive tree, the conditional POVM is a measurable operator-valued
kernel. In finite dimension, it admits a positive density with respect to
its scalar trace measure, which may depend on the preceding history. The
integrals below are understood iteratively with these conditional measures.
For a complete transcript \(y=(y_1,\ldots,y_W)\), the associated leaf effect
therefore factors as
\begin{equation}
 E_y=E^{(1)}_{y_1}\otimes E^{(2)}_{y_2|y_1}\otimes\cdots\otimes
 E^{(W)}_{y_W|y_{<W}},
 \label{eq:leaf-effect}
\end{equation}
and each factor is positive semidefinite.  Let \(\Gamma_a\) be the measurable
set of transcripts that the classical decision rule maps to action \(a\).
Coarse-graining those leaves and averaging over private randomness gives
\begin{equation}
 M_a=\int_{\Gamma_a}E_y\,d\mu(y),
 \label{eq:coarse-grained-effect}
\end{equation}
which is exactly \cref{eq:adaptive-tree-effect}.  In finite dimension, this is
a positive conic integral of product effects, equivalently a limit of finite
positive sums of product effects, and therefore belongs to the closed fully
separable cone.

Summing \cref{eq:coarse-grained-effect} over the final actions, the
sets \(\{\Gamma_a\}_a\) partition the transcript space, and recursive use
of the local completeness relations at each node gives
\(\sum_aM_a=\id_{\cH_{1:W}}\).  Finally, partial transpose on any subset of
temporal factors maps each product integrand to another positive product
operator.  Positivity is preserved by integration, so
\(M_a^{\mathsf T_S}\succeq0\) for every temporal subset \(S\), completing the
proof.
\end{proof}

\section{Memory-complexity proofs}
\label{app:memory-complexity}

For fixed \(K\), the Choi matrices of the strict update channels and the final
POVM belong to closed and bounded subsets of finite-dimensional vector spaces.
Their product is compact, and the risk is continuous; this proves attainment
in Proposition~\ref{prop:memory-monotonicity}. Embedding \(\mathbb C^K\) as a
subspace of \(\mathbb C^{K+1}\), extending each channel on the unused level,
and ensuring that the added level is never reached from the embedded input
subspace reproduces every memory-\(K\) receiver at dimension \(K+1\). The
final POVM is extended arbitrarily on the orthogonal one-dimensional
complement, for example by assigning that complement to one fixed outcome.
Because the added level is unreachable, this does not change the induced
statistics and proves monotonicity.

For Theorem~\ref{thm:full-storage}, store the first \(t\) signals coherently in
a prefix register. Immediately before the last signal arrives, its dimension
is \(d_{\mathrm{sig}}^{W-1}\). Apply the desired collective POVM when the last signal arrives,
write its outcome into one of \(A\) orthogonal terminal states, and measure that
state. A common register of dimension \(\max\{d_{\mathrm{sig}}^{W-1},A\}\) accommodates both
stages, proving the claimed sufficiency.

For the archived \(W=3,K=16\) and \(W=4,K=64\) realizations, the induced
measurement was additionally checked on a complete matrix-unit basis.
Consequently, those explicit channel sequences reproduce their target
collective measurements on arbitrary signal inputs and under extension by an
external reference system.

To prove Theorem~\ref{thm:osr}, cut the channel sequence after round
\(t\). Let \(\Phi_{1:t}\) be the composed prefix channel from
\(\cH_{1:t}\) to the \(K\)-dimensional memory. Pull the terminal effect
for action \(a\) back through the remaining channels, obtaining an operator
\(T_a\) on \(\mathbb C^K\otimes\cH_{t+1:W}\). Choose an operator basis
\(\{F_\mu\}_{\mu=1}^{K^2}\) of \(\cL(\mathbb C^K)\) and expand
\begin{equation}
 T_a=\sum_{\mu=1}^{K^2}F_\mu\otimes R_{a,\mu}^{(t)}.
\end{equation}
Pulling this operator back through the prefix channel gives
\begin{equation}
 M_a=\sum_{\mu=1}^{K^2}
 \underbrace{\Phi_{1:t}^{\dagger}(F_\mu)}_{L_\mu^{(t)}}
 \otimes R_{a,\mu}^{(t)}.
 \label{eq:osr-expansion}
\end{equation}
The rank is therefore at most \(K^2\), and the left factors depend only on
the prefix channels, not on \(a\).

\subsection*{Proof of the terminal-dimension prior bound}

Let \(F_a=\Phi^\dagger(N_a)\) be the effective input effect associated with
action \(a\), and define
\[
 u_a:=\lVert F_a\rVert_\infty .
\]
Because \(0\preceq F_a\preceq\id\), one has \(0\le u_a\le1\).  Moreover,
Lemma~\ref{lem:norm-budget} gives
\[
 \sum_a u_a\le K .
\]
Since \(S_a\succeq0\) and \(\Tr S_a=\omega_a\),
\begin{equation}
 \Tr(S_aF_a)
 \le
 \lVert F_a\rVert_\infty\Tr S_a
 =\omega_a u_a .
\end{equation}
Hence every strict memory-\(K\) receiver satisfies
\begin{align}
 P_{\rm succ}
 &=\sum_a\Tr(S_aF_a)\\
 &\le
 \max_{\substack{0\le u_a\le1\\ \sum_a u_a\le K}}
 \sum_a\omega_a u_a
 =\sum_{j=1}^{K}\omega_{(j)} .
\end{align}
Therefore
\[
 P_{K,\rm str}^{\star}(C)
 \ge
 1-\sum_{j=1}^{K}\omega_{(j)},
\]
which proves Lemma~\ref{lem:dimension-prior}.

\subsection*{Global bounds used for the two-round threshold}

The certified global lower bounds for \(K=1,2,3\) in
\cref{eq:w2-global-floors} are obtained from interval-verified feasible duals
of \cref{eq:norm-budget-dual}. Their comparison with the explicit adaptive
upper bound and the certified \(K=4\) receiver is the argument given in the
proof of Theorem~\ref{thm:w2-memory-threshold}. The minimality result uses
only the certified global lower bounds.

\section{Robustness proofs and controls}
\label{app:robustness}

For Theorem~\ref{thm:diamond-budget}, replace the ideal CPTP factors one at a
time. The telescoping hybrid expansion and diamond-norm contractivity give an
unhalved trace-norm change at most \(\sum_j\delta_j\) in the terminal classical
state. The expectation of any loss in \([0,1]\) changes by at most one half of
that distance and never by more than one, proving
Eq.~\eqref{eq:diamond-risk-bound}.

\subsection*{Certified crossing brackets}

Table~\ref{tab:crossings} gives the available dephasing-strength metadata,
in the convention \(q=1-\eta\) of
Eq.~\eqref{eq:persistent-dephasing}. The trace-preservation completion
\(\rho\mapsto\Tr(R\rho)\ket0\!\bra0\) used in the certificate is included when
the noisy curves are evaluated.

\begin{table}[h]
\centering
\small
\setlength{\tabcolsep}{3pt}
\caption{Certified dephasing-strength brackets for the fixed streaming
receivers. Exact endpoints for the two-round receivers are pending
transcription from the verification archive.}
\label{tab:crossings}
\resizebox{\columnwidth}{!}{%
\begin{tabular}{@{}lll@{}}
\toprule
Receiver & Exact bracket & Decimal\\
\midrule
\(W=2,K=4\)  & -- & \(\approx0.0169123753455\)\\
\(W=2,K=7\)  & -- & \(\approx0.0519338246576\)\\
\(W=3,K=8\)  & \(\left[\tfrac{281905}{16777216},\tfrac{281906}{16777216}\right]\)
             & \([0.01680284738,0.0168029070]\)\\
\(W=4,K=36\) & \(\left[\tfrac{5652}{16777216},\tfrac{5653}{16777216}\right]\)
             & \([0.00033688545,0.00033694506]\)\\
\bottomrule
\end{tabular}
}
\end{table}

For these particular receivers, terminal-hold dephasing applied after the last
update leaves the decision statistics unchanged: the two-round terminal POVM
effects are diagonal in the dephasing basis, and the deeper measure-and-encode
constructions produce diagonal terminal states. This is a property of these
constructions, not a general statement about processing noise.

\subsection*{Pulse-register control}

As a separate control, the incoming pulse registers were dephased before a
collective terminal measurement. At the coherent endpoint the optimized error
is approximately \(0.24210\), while complete pulse-register dephasing gives
approximately \(0.29165\). The terminal POVM is reoptimized in this control.
These values do not describe noise on the learned persistent memory and are
not used for the certified thresholds in Eq.~\eqref{eq:persistent-crossings}.

\begin{figure}[t]
\centering
\safegraphic{\columnwidth}{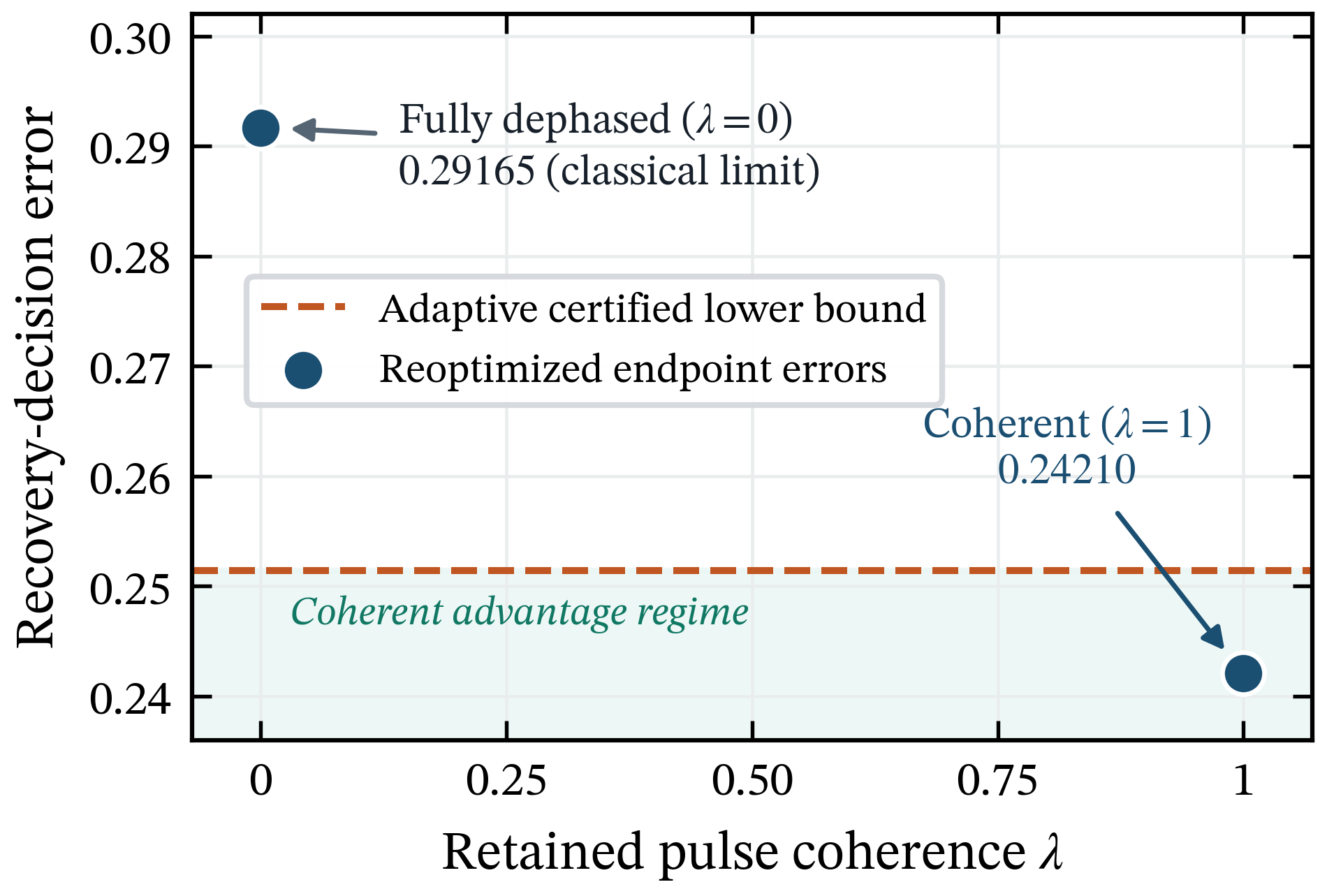}
\caption{\textbf{Pulse-register dephasing control.} The horizontal axis is
retained pulse coherence, and the vertical axis is recovery-decision error.
The points show the reoptimized endpoint errors and the dashed line the
adaptive comparison value used in this control. This auxiliary calculation
acts on stored input pulses before collective readout, not on the counted
streaming register.}
\label{fig:pulse-dephasing-control}
\end{figure}

\section{Certificate validation}
\label{app:certificate}

The certified bounds are evaluated using 256-bit Arb/Acb ball arithmetic. The
QPSK
Gram matrix, state coordinates, trajectory probabilities, and loss-weighted
cost operators are reconstructed from the analytic benchmark rather than read
as rounded cost matrices. Binary64 receiver and dual candidates are embedded
as exact rational numbers before verification, so their input conversion adds
no untracked rounding error.

For each candidate receiver, each Choi matrix and terminal effect is replaced by its
Hermitian part in ball arithmetic. After positivity of the terminal POVM
effects is verified, those effects are contracted by the exact rational factor
\(99999999/100000000\) to create completeness slack. The positive remainder is
assigned to a fixed outcome, restoring exact completeness. Channel
trace-preservation remainders are
completed by the completely positive map
\(\rho\mapsto\Tr(R\rho)\ket0\!\bra0\), with interval verification that
\(R\succeq0\). The completed maps are then checked for complete positivity
and trace preservation. The risk is evaluated from the analytic cost operators
with outward rounding, yielding an upper bound on the explicit receiver's
error. Search optimality is not required.

For the PPT dual, the exact-rational witnesses \(Y\) and \(Q_a\) are inserted
into
\begin{equation}
 Q_a\succeq0,\qquad C_a-Y-\Tpt(Q_a)\succeq0.
 \label{eq:interval-dual-check}
\end{equation}
Interval factorizations verify every cone and slack, and the objective
\(\Tr Y\) is rounded downward. Weak duality then gives a lower bound on the
analytic benchmark. For a congruence \(B=V^\dagger H V\) with invertible
\(V\), suppose a directed Gershgorin calculation gives \(B\succeq g\id\).
A lower bound on the smallest eigenvalue of \(H\) is
\begin{equation}
 \lambda_{\min}(H)\ge
 \begin{cases}
   g/\lVert V\rVert_\infty^2,&g\ge0,\\
   g\lVert V^{-1}\rVert_\infty^2,&g<0.
 \end{cases}
 \label{eq:congruence-gershgorin}
\end{equation}
The norm factors are replaced by certified upper bounds, with outward
rounding in the appropriate direction.

\begin{table}[t]
\centering
\caption{Outward-safe two-round bounds for the analytic benchmark.}
\label{tab:certificate-audit}
\resizebox{\columnwidth}{!}{%
\begin{tabular}{@{}lr@{}}
\toprule
Quantity & Certified bound\\
\midrule
\(K=7\) receiver risk & \(U_7\le0.242099929928125\)\\
Separate-adaptive risk & \(\Padstar\ge0.251431736761034\)\\
\(K=7\) separation & \(\Delta_{2,7}\ge0.009331806832908\)\\
\(K=4\) receiver risk & \(U_4\le0.249065686566581\)\\
\(K=4\) separation & \(\Delta_{2,4}\ge0.002366050194452\)\\
Collective optimum & \([0.242099896126073,0.242099929928125]\)\\
Adaptive strategy risk & \(U_{\rm ad}\le0.256727930428793\)\\
\bottomrule
\end{tabular}
}
\end{table}

The accompanying archive records the exact rational witnesses, interval
outputs, arithmetic precision, solver metadata, and cryptographic hashes used
to reproduce these directed bounds.

\section{Benchmark construction and convergence}
\label{app:benchmark}

The benchmark generator enumerates a finite latent-trajectory set
\(\cX\).  The per-round physical-error increments are
\begin{equation}
 (0,1,2,4)\quad\text{with probabilities}\quad
 \left(\tfrac{7}{10},\tfrac1{10},\tfrac1{10},\tfrac1{10}\right),
 \label{eq:increments}
\end{equation}
corresponding to no error or a single flip on one of the three physical bits.
With two rounds, at most two distinct bits can be flipped, so the reachable
final masks are \(0,1,2,3,4,5,6\); mask \(7\) is unreachable.  The eight
recovery actions are the masks \(0,\ldots,7\), and the loss is zero exactly
when the recovery mask equals the final physical-error mask.
Each record contains a fault trajectory, its probability \(\pi_x\),
the emitted syndrome labels, and the residual physical-error mask associated
with every recovery action.  If \(s_t(x)\) is the
two-bit syndrome in round \(t\), the fixed syndrome-to-QPSK lookup assigns a
phase index \(j_t(x)\in\{0,1,2,3\}\), giving
\begin{equation}
 \rho_x^{\mathrm{QPSK}}=
 \bigotimes_{t=1}^{2}
 \ket{\alpha_{j_t(x)}}\!\bra{\alpha_{j_t(x)}} .
 \label{eq:benchmark-trajectory}
\end{equation}
Temporal factors are ordered chronologically, with round \(1\) as the leftmost
tensor factor; the archive uses the same ordering.
The certified calculation constructs the finite coherent-state span directly
from the analytic Gram entries in Eq.~\eqref{eq:gram}. Arb/Acb ball arithmetic
is used for the certified cost operators and objective contractions;
tensor products of the resulting coordinates represent complete trajectories
without a Fock cutoff. In the parameter-neighborhood verification, different
values of \(\alpha\) are compared in a common analytic four-dimensional frame
using explicit isometries, rather than by subtracting unrelated Gram
factorizations.
For final physical-error mask \(b_W(x)\), the residual mask is
\(a\oplus b_W(x)\). The benchmark loss is
\begin{equation}
 \ell(a,x)=1-\delta_{a,b_W(x)},
 \label{eq:logical-loss}
\end{equation}
where the Kronecker delta is one when its indices agree and zero otherwise.
This is Eq.~\eqref{eq:mask-loss}, a comparison of complete masks rather than
logical equivalence classes.
Substitution of the enumerated \(\pi_x\), lookup \(j_t(x)\), and loss table
into \cref{eq:cost} determines every \(C_a\) without receiver-dependent
choices.

As an independent floating-point check, let \(V_\alpha^{\rm fp}\) denote the
numerical Gram factor and \(G_\alpha\) the analytic Gram matrix. The check gives
\begin{equation}
 \lVert (V_\alpha^{\rm fp})^\dagger V_\alpha^{\rm fp}-G_\alpha\rVert_F
 =2.63\times10^{-15}.
 \label{eq:gram-audit}
\end{equation}
The maximum eigenvalue discrepancy was \(2.22\times10^{-15}\).
The two-norm condition numbers are approximately
\(\kappa_2(V_\alpha^{\rm fp})=11.34\) for the Gram factor and
\(\kappa_2(G_\alpha)=128.69\) for the Gram matrix. These are independent
floating-point consistency diagnostics. The directed theorem bounds instead
use the analytic Gram construction and interval procedure of
Appendix~\ref{app:certificate}.

As an independent convergence check, the same states were also reconstructed
in a normalized truncated Fock basis,
\begin{equation}
 \ket{\alpha_j}_N=
 \frac{1}{\sqrt{Z_N}}\sum_{n=0}^{N}
 \frac{(\alpha e^{i\pi j/2})^n}{\sqrt{n!}}\ket n,\quad
 Z_N=\sum_{n=0}^{N}\frac{|\alpha|^{2n}}{n!}.
 \label{eq:fock}
\end{equation}
For each tested \(N\), the cost operators and both optimization objectives were
rebuilt.  Relative to the \(N=30\) reference, the first tested cutoff satisfying
a \(10^{-8}\) convergence criterion is \(N=8\).  At that cutoff, the absolute
differences in coherent error, repaired PPT lower bound, and gap are
\(6.22\times10^{-10}\), \(6.20\times10^{-10}\), and
\(1.87\times10^{-12}\), respectively.  This Fock sweep is a numerical
cross-check rather than an interval bound; no cutoff enters the certified
theorem.

\section{Multi-round PPT relaxation and deeper constructions}
\label{app:multi-round}

By \cref{thm:temporal-separability}, every final adaptive effect is fully
separable across temporal factors.  A tractable outer relaxation imposes
\begin{equation}
 M_a\succeq0,\qquad
 M_a^{\mathsf T_S}\succeq0
 \quad\text{for selected }S\subset\{1,\ldots,W\},
 \label{eq:multippt}
\end{equation}
together with \(\sum_aM_a=\id\).  Adding temporal bipartitions tightens the
relaxation.  Omitting a cut does not invalidate containment, but it may weaken
the resulting adaptive bound. For \(W=3\), all eight analytic residual
matrices and all twenty-four PPT multipliers were verified at 256 bits; the
scalar-repair objective debit was exactly \(0.000672565248\).

\begin{table}[h]
\centering
\caption{\textbf{Certified depth bounds and explicit streaming receivers.}
All inequalities are directed; cut metadata are stated in the text.}
\label{tab:depth}
\resizebox{\columnwidth}{!}{%
\begin{tabular}{@{}ccccc@{}}
\toprule
\(W\) & \(K\) & Receiver upper bound & PPT lower bound & Certified gap\\
\midrule
2 & 7  & 0.242099929928125 & 0.251431736761034 & 0.009331806832908\\
3 & 8  & 0.294297344448216 & 0.299647463773192 & 0.005350119324975\\
3 & 10 & 0.288350172949501 & 0.299647463773192 & 0.011297290823690\\
3 & 12 & 0.286175261363248 & 0.299647463773192 & 0.013472202409943\\
3 & 16 & 0.284502699489212 & 0.299647463773192 & 0.015144764283980\\
4 & 36 & 0.319370057047514 & 0.319542490402679 & 0.000172433355164\\
4 & 48 & 0.317851443330621 & 0.319542490402679 & 0.001691047072057\\
4 & 64 & 0.317529609881821 & 0.319542490402679 & 0.002012880520857\\
\bottomrule
\end{tabular}
}
\end{table}

The \(W=3,K=16\) and \(W=4,K=64\) entries are explicit full-storage
streaming realizations constructed using Theorem~\ref{thm:full-storage}. Their
Choi matrices and terminal POVMs are materialized exactly in sparse
rational and Gaussian-rational form; the \(W=4\) matrices are represented
sparsely.

The remaining entries are lower-memory streaming receivers obtained from a
restricted prefix-memory construction. Selected prefix
computational-basis vectors are preserved coherently, the discarded population
is reset to \(\ket0\), and the selection keeps the vectors of largest average
diagonal mass with deterministic tie-breaking. Only the terminal
discrimination POVM is then numerically optimized and certified. Because the
construction is restricted, a failure at some \(K\) does not exclude that
dimension globally. The tested dimensions were
\(K=4,5,6,7,8,9,10,11,12,16\) at \(W=3\) and
\(K=4,8,12,16,24,32,34,35,36,40,48,64\) at \(W=4\). For \(K<8\) at \(W=3\)
this measure-and-encode family supports at most \(K\) terminal labels, so the
negative results there are constructional rather than global; this is why the
\(W=3\) statement remains \(3\le K_{\rm sep}\le8\).

For \(W=3\), the relaxation uses the cuts \(1|23\), \(2|13\),
and \(3|12\); for \(W=4\), it uses \(12|34\). In particular, the \(W=4\)
certificate does not establish an advantage for a constrained \(K=4\)
receiver.

\section{A separable-optimum example}
\label{app:sep-example}

Separability across the temporal cut does not by itself preclude a coherent
advantage. Let \(b\in\{0,1\}\) and \(\beta\in\{Z,X\}\) be independent and
uniform. The first signal is
\begin{equation}
\begin{aligned}
 \ket{\psi_{0,Z}}&=\ket0,&
 \ket{\psi_{1,Z}}&=\ket1,\\
 \ket{\psi_{0,X}}&=\ket+,&
 \ket{\psi_{1,X}}&=\ket-,
\end{aligned}
 \label{eq:sep-example-states}
\end{equation}
where \(\ket\pm=(\ket0\pm\ket1)/\sqrt2\), and \(X,Z\) label their
measurement bases. The second signal is an orthogonal basis flag \(\ket\beta\). Thus
\begin{equation}
 \rho_{b,\beta}
 =\ket{\psi_{b,\beta}}\!\bra{\psi_{b,\beta}}
 \otimes\ket{\beta}\!\bra{\beta},
 \label{eq:sep-example-source}
\end{equation}
and the final action is a guess \(a\in\{0,1\}\) with
\(\ell(a,b,\beta)=\mathbf 1[a\ne b]\).

A two-dimensional coherent memory can retain the first qubit until the basis
flag arrives and then measure in the indicated basis, giving zero error. A
receiver that must measure the first signal before learning the basis has
optimal success probability \(\cos^2(\pi/8)=(1+1/\sqrt2)/2\)
\citep{ballester2008postmeasurement}, and hence error \((1-1/\sqrt2)/2\).
Therefore
\begin{equation}
\begin{aligned}
 P_{2,\rm str}^{\star}&=\Pcollstar=0,\\
 \Padstar&=\frac{1-1/\sqrt2}{2},\\
 K_{\rm sep}&=K_{\rm coll}=2 .
\end{aligned}
 \label{eq:sep-example}
\end{equation}
The optimal collective measurement is nevertheless separable across the
temporal cut,
\begin{equation}
 M_b=
 \ket{\psi_{b,Z}}\!\bra{\psi_{b,Z}}\otimes\ket Z\!\bra Z
 +\ket{\psi_{b,X}}\!\bra{\psi_{b,X}}\otimes\ket X\!\bra X .
 \label{eq:sep-example-povm}
\end{equation}
The example shows that a PPT or separability relaxation cannot detect every
temporal-order advantage, and it is not used in any certificate above.

\section{Observation-only depth control}
\label{app:observation-control}

As a separate numerical control, additional observations were introduced
without additional data-error opportunities.  This differs from both the
certified two-round benchmark and the slow timing schedule used for
Fig.~\ref{fig:hardware-t2}.  Within that control model, the coherent
errors decrease from approximately \(8.761\%\) to \(4.093\%\) to \(1.814\%\)
for two, three, and four observations.  The corresponding
dephasing-threshold parameters are
\(\lambda_\star^{(2)}=0.9402\),
\(\lambda_\star^{(3)}=0.9863\), and
\(\lambda_\star^{(4)}=0.9966\).
Here \(\lambda_\star\) denotes the coherence-retention parameter used in this
control calculation.
These values pertain only to the observation-only control and are not used in
the stored-pulse hardware calculation.

\section{Numerical optimization of streaming updates}
\label{app:choi}

Consider a channel
\(\cE:\cL(\mathbb C^{d_{\mathrm{in}}})\to
\cL(\mathbb C^{d_{\mathrm{out}}})\),
where \(d_{\mathrm{in}}\) and \(d_{\mathrm{out}}\) are the input and
output dimensions. Let \(\{\ket i\}\) be an orthonormal input basis.
Its Choi matrix is defined by

\begin{equation}
 J(\cE)=\sum_{i,j=1}^{d_{\mathrm{in}}}
 \ket i\!\bra j\otimes\cE(\ket i\!\bra j).
 \label{eq:choi}
\end{equation}
Complete positivity and trace preservation are equivalent to
\begin{equation}
 J(\cE)\succeq0,\qquad
 \Tr_{\mathrm{out}}J(\cE)=\id_{\mathrm{in}}.
 \label{eq:cptp}
\end{equation}
The action of the channel is recovered from
\begin{equation}
 \cE(X)=\Tr_{\mathrm{in}}\!\left[
 (X^{\mathsf T}\otimes\id_{\mathrm{out}})J(\cE)\right].
 \label{eq:choiaction}
\end{equation}
After composing the update maps with the loss-weighted input ensemble, the
average risk is multilinear in the channel Choi matrices and linear in the
final POVM.  Holding every block but one fixed therefore leaves a linear
objective over the positive semidefinite and trace-preserving constraints in
\cref{eq:cptp}; optimizing that block is an SDP.  The final POVM block is
likewise an SDP with positivity and completeness constraints.

The streaming search uses block-coordinate optimization.  It begins from
feasible channel and POVM blocks, replaces one block at a time by the solution
of its convex subproblem, and repeats channel and measurement updates until the
prescribed stopping criterion is reached.  An exactly solved block update cannot
increase the objective, but finite solver tolerances can obscure changes at the
last reported digits. Figure~\ref{fig:memory} shows the search values;
Table~\ref{tab:memory} gives the certified bounds used for minimality.

Feasibility can be checked independently of the optimization expression.  The
candidate receivers are evaluated by reconstructing the induced channel action
with \cref{eq:choiaction} and recomputing the recovery risk from the resulting
output states. The feasibility check records the smallest Choi and POVM
eigenvalues, the trace-preservation residuals, and the Frobenius-norm POVM
completeness residual \(\lVert\sum_aN_a-\id\rVert_F\). These diagnostics are
used to screen the candidate receivers before interval verification.

For the \(K=7\) and \(K=4\) receivers, the feasibility check found
trace-preservation residuals below \(9.7\times10^{-15}\) and Choi minimum
eigenvalues no more than \(7\times10^{-17}\) below zero.  The final POVMs were
positive semidefinite and complete to floating-point precision.  Before
interval verification, the repaired floating-point candidates had risks
\(0.242099914992664\) for \(K=7\) and \(0.249065671918806\) for \(K=4\). The
certified theorem instead uses the outward-safe upper bounds in
Table~\ref{tab:certificate-audit}.

Block-coordinate optimization therefore supplies candidate receivers but does
not establish their global optimality.  Feasibility and outward-safe risk
bounds for the receivers used in the certified results are obtained
independently by the procedure of Appendix~\ref{app:certificate}. The
reproducibility archive records the initialization rules, block order, stopping
tolerance, completed starts, solver versions, and feasibility residuals for
the reported memory sweep.

\section{Clocked quantum-reservoir pilot}
\label{app:qrc}

We compare homogeneous and clocked reservoir constructions on the \(W=3\)
pilot benchmark with reservoir memory dimension \(m=4\), where \(m\) is the
dimension of the persistent reservoir register carried between rounds.  The
clocked and parameter-matched homogeneous families were compared over thirty
seeds, grouped into ten matched best-of-three trials.
Clocking gave lower error in 29 of 30 raw seed comparisons and in 10 of 10
grouped trials, with a mean difference of \(0.00200414101285111\) in absolute
error, that is \(0.200414101285111\) percentage points.  Because the
comparison was not designed as a prespecified inferential study, we report
these results descriptively.

\begin{table}[t]
\centering
\caption{Descriptive clocked-reservoir comparison. No inferential interval is
reported.}
\label{tab:qrc-paired}
\resizebox{\columnwidth}{!}{%
\begin{tabular}{@{}lr@{}}
\toprule
Quantity & Reported result\\
\midrule
Raw seed comparisons & 29/30 clocked wins\\
Grouped best-of-three trials & 10/10 clocked wins\\
Mean error reduction & 0.200414101285111 percentage points\\
\bottomrule
\end{tabular}
}
\end{table}

Within this guided comparison, the clocked family gives lower error than the
matched homogeneous family. Because the clocked maps were informed by the
optimized coherent receiver, we treat this comparison as exploratory.  The per-seed results and the grouping rule
are included in the archive; the original training source and checkpoints are
not.

\section{Hardware-noise sensitivity}
\label{app:hardware-sensitivity}

\begin{figure}[t]
\centering
\safegraphic{\columnwidth}{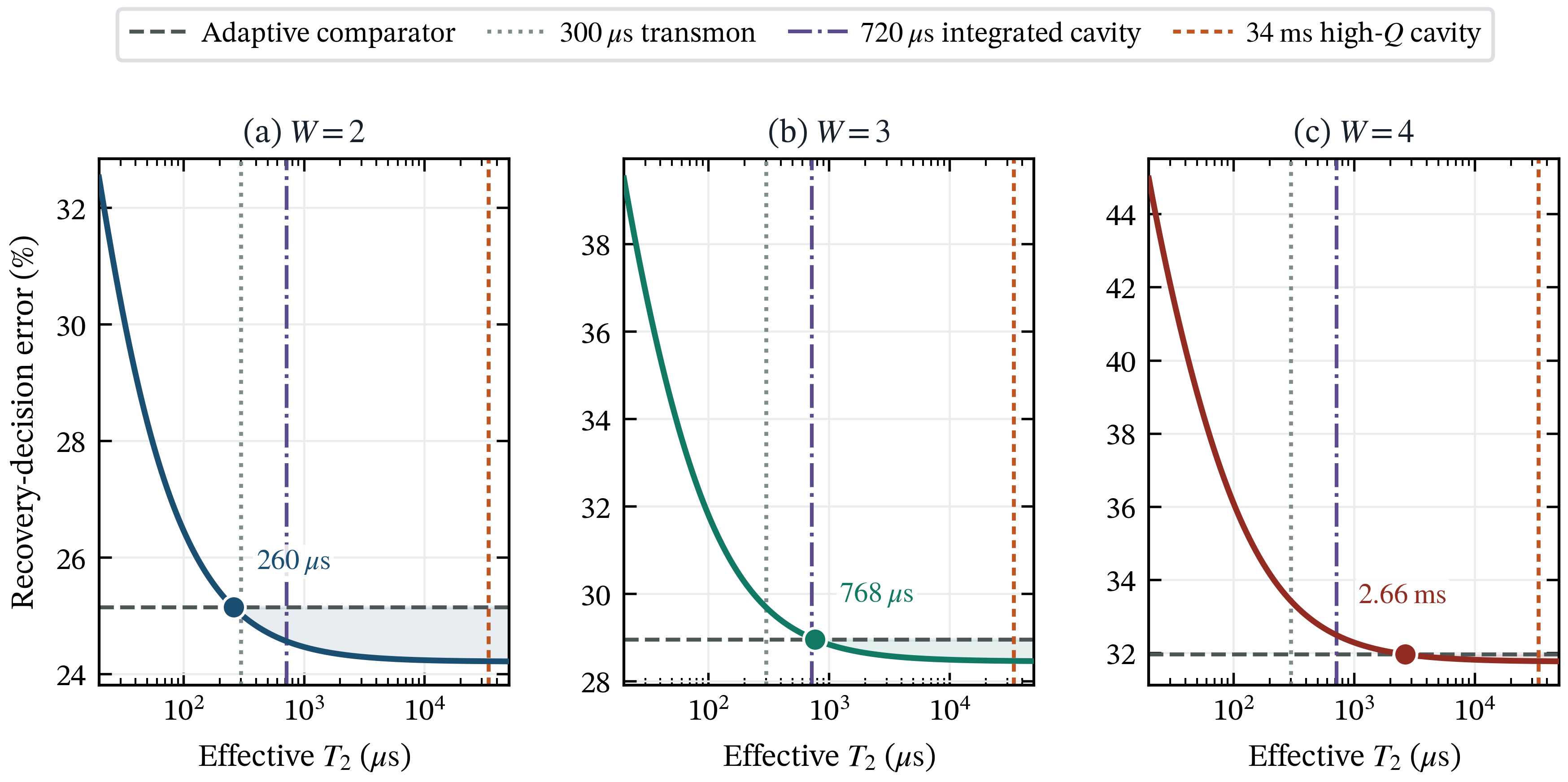}
\caption{\textbf{Stored-pulse collective-readout model.} Solid curves show
the slow-schedule calculation in which the final joint POVM is reoptimized at
each \(T_2\). Each panel has coherence time on the horizontal axis and
recovery-decision error on the vertical axis; panels correspond to \(W=2,3,4\).
Horizontal dashed lines show the comparator values used in the
stored-pulse root calculation; the \(W=3\) root uses the single-cut comparator
employed in that stored-pulse calculation rather than the three-cut bound in
Eq.~\eqref{eq:w3-cuts}. The
marked crossings give \(T_{2,\star}=259.942\,\mu\mathrm{s}\),
\(767.970\,\mu\mathrm{s}\), and \(2656.778\,\mu\mathrm{s}\) for \(W=2,3,4\).
Vertical lines indicate separately reported component-scale coherence values:
approximately \(300\,\mu\mathrm{s}\) for a tantalum transmon
\citep{place2021transmon}, \(720\,\mu\mathrm{s}\) for an integrated
cavity memory \citep{reagor2016memory}, and \(34\,\mathrm{ms}\) for a high-Q
cavity qubit \citep{milul2023cavity}.  These coherence values were reported in
separate experiments and do not represent an integrated realization of the
receiver considered here.}
\label{fig:hardware-t2}
\end{figure}

The modeled round duration is
\begin{equation}
 \tau_{\rm round}=(5.75+0.05+0.02)\,\mu\mathrm{s}=5.82\,\mu\mathrm{s},
 \label{eq:tau-round-decomposition}
\end{equation}
combining a separately simulated pulse duration with assumed
memory-interaction and inter-round gap times. All three components are
modeling assumptions rather than measurements of the receiver studied here.
The three crossings in Fig.~\ref{fig:hardware-t2} were located by Brent root
finding in \(\log T_2\), with an unrestricted joint POVM reoptimized at each
trial point.

For context, superconducting microwave-cavity memories have demonstrated
long-lived storage \citep{reagor2016memory}, tantalum transmons have exceeded
\(0.3\,\mathrm{ms}\) under dynamical decoupling \citep{place2021transmon},
and a single-photon cavity qubit reached
\(T_2^c=34\pm1\,\mathrm{ms}\) \citep{milul2023cavity}. Separate experiments
have demonstrated \(140\,\mathrm{ns}\) superconducting-qubit readout
\citep{jeffrey2014fast} and a \(2.64\,\mu\mathrm{s}\)
measurement-to-conditional-pulse sequence \citep{riste2012feedback}. These
records were not achieved together as a coherent sequential receiver, and the
\(140\,\mathrm{ns}\) result characterizes a fast measurement step rather than
coherent capture, storage, and reinteraction. Because maximum latency depends
on whether pulse integration, control, and memory interaction overlap, we
report lifetime values rather than a single architecture-independent latency
ceiling.

\begin{table}[h]
\centering
\small
\setlength{\tabcolsep}{3pt}
\caption{\textbf{Stored-pulse slow-schedule crossing values.} The values are
model-dependent and do not represent integrated-receiver performance.}
\label{tab:hardware}
\begin{tabular}{@{}cccc@{}}
\toprule
Depth & Longest storage time \(t_1\) & \(\lambda_\star\) & \(T_{2,\star}\)\\
\midrule
\(W=2\) & \(8.46\,\mu\mathrm{s}\) & \(0.967978\) & \(260\,\mu\mathrm{s}\)\\
\(W=3\) & \(14.28\,\mu\mathrm{s}\) & \(0.981577\) & \(768\,\mu\mathrm{s}\)\\
\(W=4\) & \(20.10\,\mu\mathrm{s}\) & \(0.992463\) & \(2.66\,\mathrm{ms}\)\\
\bottomrule
\end{tabular}
\end{table}

Equation~\ref{eq:t2} models phase retention, whereas amplitude damping is
controlled separately by \cref{eq:t1}.  Under a Markovian decomposition,
\cref{eq:t1t2} separates the pure-dephasing time \(T_\phi\) from the
amplitude-relaxation time \(T_1\).  This calculation includes phase retention
only; it is not a complete device-noise model. A
multilevel memory also requires a specified encoding and decay law
for its off-diagonal matrix elements; the cited qubit coherence times alone
do not determine those rates.
A device-level assessment would vary \(T_1\) and \(T_\phi\) independently and
propagate timing and control uncertainties.

\section{Flagged direct-sum inheritance criterion}
\label{app:flagged}

A certified base-sector separation can be inherited by a larger
block-diagonal decision problem under the following conditions.

\begin{proposition}[Locally flagged inheritance]
\label{prop:flagged}
Let an orthogonal classical flag be accessible to both receiver classes
before any operation whose choice depends on that flag. Assume the enlarged
receiver classes admit the causal embedding of the base-sector strategies,
and that restriction of an enlarged adaptive strategy to the base sector
is an allowed base adaptive strategy.  Let its register \(F\) belong wholly to one local temporal tensor
factor, and require that no partial transpose used by the certificate splits
\(F\) from that factor.  In the fixed flag basis define the pinching map
\begin{equation}
 \cP_F(X):=\sum_f
 (\ket{f}\!\bra{f}_F\otimes\id)X
 (\ket{f}\!\bra{f}_F\otimes\id).
 \label{eq:flag-pinching}
\end{equation}
Suppose the enlarged cost operators are block diagonal,
\begin{equation}
 \widehat C_a=q C_a^{(0)}\oplus(1-q)D,
 \qquad 0<q\le1,
 \label{eq:flagged-cost}
\end{equation}
where \(D\) is independent of the action.  If the base sector has an adaptive
lower bound \(L_{\adapt}\), a fixed receiver \(M^{\rm coh}\) with explicit
risk \(r(M^{\rm coh};C^{(0)})\), and certified gap
\(\delta=L_{\adapt}-r(M^{\rm coh};C^{(0)})>0\), then the enlarged problem has the
explicit-versus-certified separation
\begin{equation}
 \widehat\Delta_{\rm cert}\ge q\delta.
 \label{eq:flagged-gap}
\end{equation}
If the conditional base-sector costs are perturbed by \(\Delta C_a\), the
right-hand side becomes
\(q[\delta-2\epsilon_*]\), with \(\epsilon_*\) from
\cref{eq:epsilon-star}.
\end{proposition}

\begin{proof}
Pinching a feasible POVM in the orthogonal flag basis preserves positivity,
completeness, and the block-diagonal objective. Since the flag belongs to
one temporal factor and its projectors are diagonal in the transpose basis,
for every certified cut \(S\),
\begin{equation}
 \cP_F\circ\mathsf T_S=\mathsf T_S\circ\cP_F.
 \label{eq:flag-commutes}
\end{equation}
Thus pinching also preserves the PPT relaxation and permits sectorwise
analysis. The lower bound for the adaptive class uses a separate assumption:
restricting an enlarged adaptive protocol to the base sector must give an
allowed base-sector adaptive protocol. Its conditional risk is therefore at
least \(L_{\adapt}\). Conversely, the causal-embedding assumption lets the
coherent receiver use \(M^{\rm coh}\) on the base sector. In the other sector,
the action-independent cost gives \(\sum_a\Tr(DM_a)=\Tr D\), irrespective
of the receiver. This common contribution cancels, leaving a gap of at least
\(q[L_{\adapt}-r(M^{\rm coh};C^{(0)})]\). The perturbed statement follows by
applying Proposition~\ref{prop:stability} to the base-sector costs.
\end{proof}

The same reasoning applies when the complementary sector is not
action-independent only if a nonnegative explicit-versus-certified
contribution is established there separately.  The inheritance therefore
requires the flag, cost decomposition, and relaxation to respect the same
direct-sum structure.  For a family indexed by code
distance \(d_{\mathrm{code}}\), a uniform inherited floor would require both
\(q_{d_{\mathrm{code}}}\ge q_0>0\) and
\(\epsilon_{*,d_{\mathrm{code}}}\le\epsilon_0<\delta/2\), giving
\begin{equation}
 \widehat\Delta_{{\rm cert},d_{\mathrm{code}}}
 \ge q_0(\delta-2\epsilon_0)>0 .
 \label{eq:flagged-family}
\end{equation}
Neither condition has been established for a natural surface-code family; in
particular, the probability of a chosen sector may vanish with distance.  Thus
the proposition is conditional and does not establish scaling for a natural
surface-code family.


\begin{acknowledgments}
\noindent This work was authored in part by the National Laboratory of the Rockies (NLR) for the U.S. Department of Energy (DOE), operated under Contract No. DE-AC36-08GO28308. Funding for the NLR authors provided by the Modeling, Simulation, Optimization Capability for Grid Modernization, Energy Systems Integration, Mobility Electrification (MSOC 4 GEM) at the National Laboratory of the Rockies. The views expressed in the article do not necessarily represent the views of the DOE or the U.S. Government. The U.S. Government retains and the publisher, by accepting the article for publication, acknowledges that the U.S. Government retains a nonexclusive, paid-up, irrevocable, worldwide license to publish or reproduce the published form of this work, or allow others to do so, for U.S. Government purposes. Grammarly was used for grammar checking during the preparation of this manuscript. The authors reviewed and edited the suggested changes and take full responsibility for the final content. 


\end{acknowledgments}

\bibliographystyle{apsrev4-2}
\bibliography{references}

\end{document}